\documentclass[fleqn,usenatbib]{mnras}

\usepackage{newtxtext,newtxmath}

\usepackage[T1]{fontenc}

\DeclareRobustCommand{\VAN}[3]{#2}
\let\VANthebibliography\thebibliography
\def\thebibliography{\DeclareRobustCommand{\VAN}[3]{##3}\VANthebibliography}

\usepackage{graphicx,animate}	
\usepackage{amsmath}	
\usepackage{ragged2e}
\usepackage[encapsulated]{CJK}

\title[UNCOVER lens model for Abell~2744]{UNCOVERing the extended strong lensing structures of Abell~2744 with the deepest JWST imaging}

\author[L. J. Furtak et al.]{Lukas J. Furtak,$^{1}$\thanks{E-mail: furtak@post.bgu.ac.il}
Adi Zitrin,$^{1}$
John R. Weaver,$^{2}$
Hakim Atek,$^{3}$
Rachel Bezanson,$^{4}$
Ivo Labb\'{e},$^{5}$
\newauthor Katherine E. Whitaker,$^{2}$
Joel Leja,$^{6,7,8}$
Sedona H. Price,$^{4}$
Gabriel B. Brammer,$^{9}$
Bingjie Wang (\begin{CJK*}{UTF8}{gbsn}王冰洁\ignorespacesafterend\end{CJK*}),$^{6,7,8}$
\newauthor Danilo Marchesini,$^{10}$
Richard Pan,$^{10}$
Pratika Dayal,$^{11}$
Pieter van Dokkum,$^{12}$
Robert Feldmann,$^{13}$
\newauthor Seiji Fujimoto,$^{14,9}$
Marijn Franx,$^{15}$
Gourav Khullar,$^{4}$
Erica J. Nelson$^{16}$ 
and Lamiya A. Mowla$^{17}$
\\
$^{1}$Physics Department, Ben-Gurion University of the Negev, P.O. Box 653, Be’er-Sheva 84105, Israel\\
$^{2}$Department of Astronomy, University of Massachusetts, Amherst, MA 01003, USA\\
$^{3}$Institut d'Astrophysique de Paris, CNRS, Sorbonne Universit\'e, 98bis Boulevard Arago, 75014, Paris, France\\
$^{4}$Department of Physics and Astronomy and PITT PACC, University of Pittsburgh, Pittsburgh, PA 15260, USA\\
$^{5}$Centre for Astrophysics and Supercomputing, Swinburne University of Technology, Melbourne, VIC 3122, Australia\\
$^{6}$Department of Astronomy \& Astrophysics, The Pennsylvania State University, University Park, PA 16802, USA\\
$^{7}$Institute for Computational \& Data Sciences, The Pennsylvania State University, University Park, PA 16802, USA\\
$^{8}$Institute for Gravitation and the Cosmos, The Pennsylvania State University, University Park, PA 16802, USA\\
$^{9}$Cosmic Dawn Center (DAWN), Niels Bohr Institute, University of Copenhagen, Jagtvej 128, K{\o}benhavn N, DK-2200, Denmark\\
$^{10}$Department of Physics and Astronomy, Tufts University, 574 Boston Ave., Medford, MA 02155, USA\\
$^{11}$Kapteyn Astronomical Institute, University of Groningen, P.O. Box 800, 9700 AV Groningen, The Netherlands\\
$^{12}$Department of Astronomy, Yale University, New Haven, CT 06511, USA\\
$^{13}$Institute for Computational Science, University of Zurich, Winterhurerstrasse 190, CH-8006 Zurich, Switzerland\\
$^{14}$Department of Astronomy, The University of Texas at Austin, Austin, TX 78712, USA\\
$^{15}$Leiden Observatory, Leiden University, P.O. Box 9513, 2100 RA Leiden, Netherlands\\
$^{16}$Department for Astrophysical and Planetary Science, University of Colorado, Boulder, CO 80309, USA\\
$^{17}$Dunlap Institute for Astronomy and Astrophysics, 50 St. George Street, Toronto, Ontario, M5S 3H4, Canada\\
}

\date{Accepted 2023 May 25. Received 2023 May 25; in original form 2022 December 16}

\pubyear{2022}

\begin{document}
\label{firstpage}
\pagerange{\pageref{firstpage}--\pageref{lastpage}}
\maketitle

\begin{abstract}
We present a new parametric lens model for the massive galaxy cluster Abell~2744 based on new ultra-deep JWST imaging taken in the framework of the UNCOVER program. These observations constitute the deepest JWST images of a lensing cluster to date, adding to existing deep \textit{Hubble Space Telescope} (HST) images and the recent JWST ERS and DDT data taken for this field. The wide field-of-view of UNCOVER ($\sim45$\,arcmin$^2$) extends beyond the cluster's well-studied central core and reveals a spectacular wealth of prominent lensed features around two massive cluster sub-structures in the north and north-west, where no multiple images were previously known. We identify 75 new multiple images and candidates of 17 sources, 43 of which allow us, for the first time, to constrain the lensing properties and total mass distribution around these extended cluster structures using strong lensing (SL). Our model yields an effective Einstein radius of $\theta_{E,\mathrm{main}}=23.2\arcsec\pm2.3\arcsec$ for the main cluster core (for $z_{\mathrm{s}}=2$), enclosing a mass of $M(<\theta_{E,\mathrm{main}})=(7.7\pm1.1)\times10^{13}$\,M$_{\odot}$, and $\theta_{E,\mathrm{NW}}=13.1\arcsec\pm1.3\arcsec$ for the newly discovered north-western SL structure enclosing $M(<\theta_{E,\mathrm{NW}})=(2.2\pm0.3)\times10^{13}$\,M$_{\odot}$. The northern clump is somewhat less massive with $\theta_{E,\mathrm{N}}=7.4\arcsec\pm0.7\arcsec$ enclosing $M(<\theta_{E,\mathrm{N}})=(0.8\pm0.1)\times10^{13}$\,M$_{\odot}$. We find the northern sub-structures of Abell~2744 to broadly agree with the findings from weak lensing (WL) analyses and align with the filamentary structure found by these previous studies. Our model in particular reveals a large area of high magnification values between the various cluster structures, which will be paramount for lensed galaxy studies in the UNCOVER field. The model is made publicly available to accompany the first UNCOVER data release.
\end{abstract}




\section{Introduction} \label{sec:intro}
Galaxy clusters are the most massive gravitationally bound structures in the Universe. As such, they form relatively late in the hierarchical cosmic history, around $z\lesssim1-3$ \citep[e.g.][]{amoura21}. Clusters form at the nodes of the cosmic web \citep[e.g.][]{zeldovich82,delapparent86} where streams of matter and galaxies fall in towards the dense cluster cores, as shown in detailed N-body simulations \citep[e.g.][]{klypin83,springel05,dubois14,hatfield19,ata22}. Indeed, massive galaxy clusters often show merging sub-structures in their central regions, along with disturbed X-ray gas or intra-cluster light (ICL) morphologies, whereas some also show prominent cluster structures far outside their core \citep[e.g.][]{borgani01,ebeling04,kartaltepe08,jauzac12,jauzac16b,jauzac18,medezinski16,durret16,ellien19,kim19,asencio21,diego22}. The existence, and the observed number of the most massive galaxy clusters can be confronted with expectations from $\Lambda$CDM cosmology \citep[e.g.][]{mullis05,ebeling07,foley11,stanford12,wang16,miller18,asencio21} and therefore help constrain the formation and evolution of large-scale structures in the Universe.

One of the most direct methods to detect massive clusters and constrain their total matter distributions is through their gravitational lensing signatures. Although the triaxial shape of clusters creates some bias towards clusters that are elongated along the light of sight \citep[e.g.][]{hennawi07,sereno10,okabe10,oguri12}, strong lensing (SL) clusters showing a large number of multiply imaged features are often also the most massive galaxy clusters. These are usually either more concentrated, with higher masses in their central core \citep[e.g.][]{broadhurst08a,broadhurst08b}, or yet un-virialized merging systems with multiple cores and dark matter (DM) sub-structures enhancing the SL regime \citep[e.g.][]{meneghetti03,meneghetti07a,meneghetti07b,torri04,redlich12,zitrin13a}. Gravitational lensing, and in particular the SL regime, thus represents a unique means to probe the otherwise invisible DM distribution in massive galaxy clusters. 

Galaxy clusters also act as `cosmic telescopes': The gravitational magnification of background objects allows us to detect and study high-redshift galaxies that would otherwise be too faint to be observed \citep[e.g.][]{maizy10,kneib11,sharon12,zitrin14,monna14,richard14,coe15,coe19}. Indeed, several observational campaigns targeting galaxy clusters undertaken over the past decade with the \textit{Hubble Space Telescope} (HST), often complemented by ground-based observations, have led to the detection of several hundreds of lensed high-redshift galaxies at $z>6$, down to UV luminosities of $M_{\mathrm{UV}}\lesssim-13$ \citep[e.g.][]{livermore17,bouwens17,bouwens22a,ishigaki18,atek18} and stellar masses $M_{\star}\gtrsim10^6\,\mathrm{M}_{\odot}$ \citep{bhatawdekar19,kikuchihara20,furtak21,strait20,strait21}. These campaigns to observe SL clusters have also enabled the spectacular detections of transient phenomena such as lensed supernovae \citep[e.g.][]{kelly15,kelly16a,kelly16b,rodney15,rodney22} and even single stars \citep[e.g.][]{kelly18,welch22,meena22a}.

Abell~2744 \citep[e.g.][]{abell89,allen98,ebeling10} at $z_{\mathrm{d}}=0.308$ has by now become a well-studied galaxy cluster. Largely owing to its lensing strength, it was chosen for the \textit{Hubble Frontier Fields} \citep[HFF;][]{lotz17} program, an HST Director's Discretionary Time (DDT) program that targeted six lensing clusters to unprecedented depth. Abell~2744 has also been extensively targeted by other ground- and space-based follow-up programs. In terms of SL analyses, the main cluster core was first analyzed in \citet{merten11} using the \texttt{Light-Traces-Mass} modeling method \citep[\texttt{LTM};][]{zitrin09a}, revealing a large number of multiply imaged systems. \citet{merten11} also used X-ray data and performed a wider-field weak lensing (WL) analysis, finding other massive structures around the main clump, which already hinted at the merging history of the cluster. Several lens modeling teams have since published SL models for the core of Abell~2744 based on HFF imaging and spectroscopic follow-up campaigns \citep{richard14,johnson14,wang15,jauzac15,kawamata16,kawamata18,priewe17}. These also yielded numerous lensed high-redshift galaxies behind Abell~2744 \citep[e.g.][]{zheng14,atek14a,atek15a,ishigaki15a}, including a triply imaged $z\sim10$ galaxy \citep{zitrin14}, recently confirmed spectroscopically by \citet{roberts-borsani22}, and a multiply imaged candidate system at $z\sim7-8$ \citep[][]{atek14a,mahler18}, which we confirm is a triply-imaged high-redshift object. This particular object seems to be a uniquely red and compact high-redshift object, potentially quasar-like, as presented in detail in a dedicated paper \citet{furtak22b}. The detailed WL studies in a wide field around the core revealed substantial extended DM sub-structures to the far north and north-west of the cluster core, which seem to hint at a large-scale filamentary structure \citep{medezinski16,jauzac16b}. The most recent SL models of the cluster were published by \citet{mahler18}, refined in \citet{richard21}, and by \citet{bergamini22}. The \citet{mahler18} model is constrained by 188 multiple images, many of which are photometric, and the \citet{bergamini22} model includes 90 spectroscopically confirmed multiple images in the main cluster center. Both of these state-of-the art models included DM halo components far ($\sim3\arcmin$) outside the main cluster core. These were included to account for the substructures discovered in previous WL analyses, and following galaxy over-densities seen in wider-field HST imaging around the cluster. However, due to the lack of ultra-deep high-resolution imaging, and thus SL constraints far from the cluster center, these external DM halos were not directly constrained in these models\footnote{Though see our latest HST-based model used in \citet{roberts-borsani22} where we included three tentative multiple systems identified around the two northern clumps in archival HST data in a first attempt to constrain these sub-structures.}.

This now changes thanks to the unprecedented sensitivity and spatial resolution of the JWST \citep{gardner06,mcelwain23}, the advent of which has already revealed numerous new SL features in several clusters and enabled significant improvements to their SL models \citep{mahler22,pascale22,caminha22,diego22,hsiao22,williams22,meena22b}. Most recently, the JWST targeted Abell~2744 in the framework of the \textit{Ultra-deep NIRSpec and NIRCam ObserVations before the Epoch of Reionization} (UNCOVER) program 
\citep[PIs: I. Labb\'{e} \& R. Bezanson;][]{bezanson22} for wide-field ultra-deep imaging of the cluster and its surroundings (it was also observed to lesser depth in two other JWST programs; see section~\ref{sec:data}). These new observations now reveal spectacular multiple imaging and strongly lensed arcs also outside the core of Abell~2744, namely around the northern and most prominently the north-western sub-structures. In this work, we use the UNCOVER imaging to identify and classify these new SL constraints and for the first time directly constrain the northern and north-western extensions of the cluster, with a total of 43 new multiple images. These are then incorporated into a new parametric SL model containing the entire central UNCOVER field-of-view, covering a total area of about $7.6\arcmin\times7.6\arcmin$. The lensing maps of our model are made publicly available as part of the UNCOVER data product release\footnote{The maps can be found in the following repository: \url{https://jwst-uncover.github.io/DR1.html##LensingMaps}, which is part of the UNCOVER website \url{https://jwst-uncover.github.io}.}.

This paper is organized as follows: We briefly present the targeted cluster and the data used in this analysis in section~\ref{sec:data}, before describing our modeling methods and the newly identified lensed features used to constrain our SL model in section~\ref{sec:modeling}. We present and discuss the results in section~\ref{sec:result} and the work is concluded in section~\ref{sec:conculsion}. Throughout this paper, we assume a standard flat $\Lambda$CDM cosmology with $H_0=70\,\frac{\mathrm{km}}{\mathrm{s}\,\mathrm{Mpc}}$, $\Omega_{\Lambda}=0.7$, and $\Omega_\mathrm{m}=0.3$. All magnitudes quoted are in the AB system \citep{oke83} and all quoted uncertainties represent $1\sigma$ ranges unless stated otherwise.

\begin{figure*}
    \centering
    \includegraphics[width=\textwidth, keepaspectratio=true]{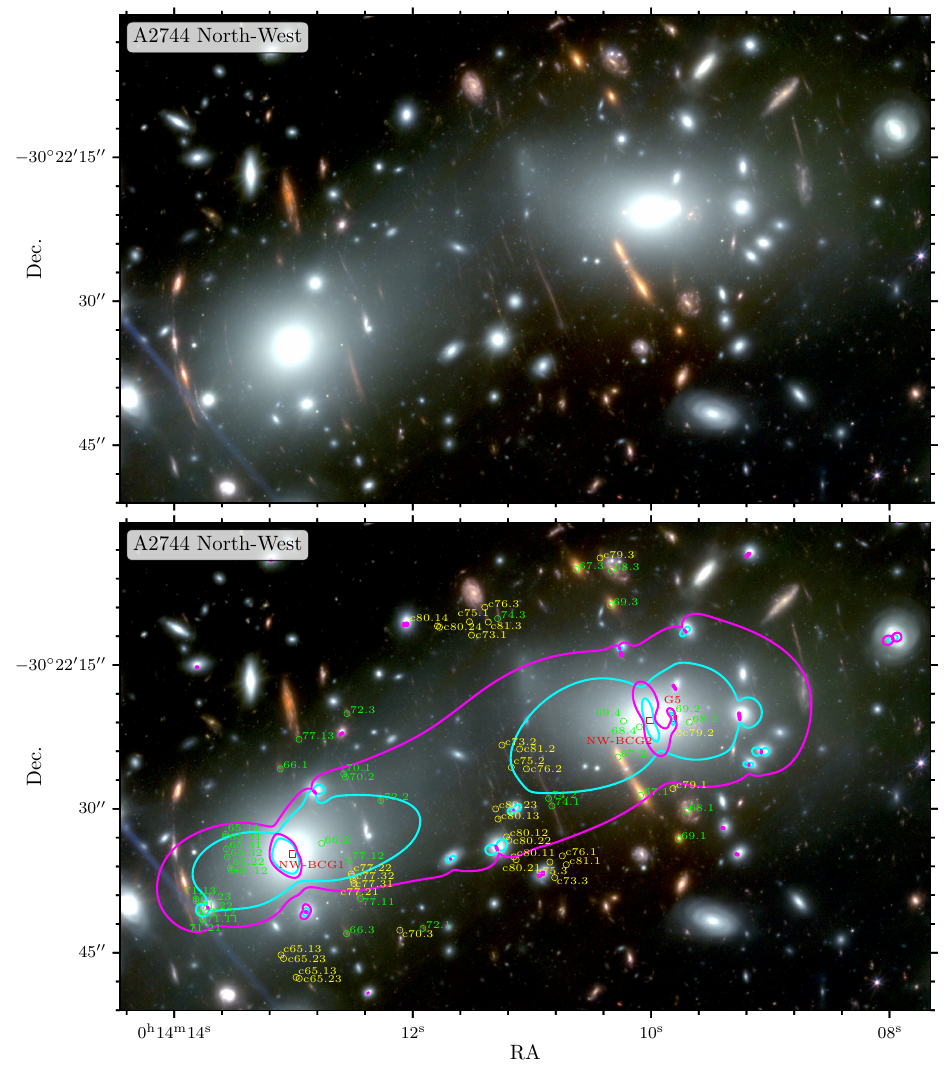}
    \caption{A $1.46\arcmin\times0.85\arcmin$ cutout of an UNCOVER color-composite image exhibiting the many newly discovered SL features and candidates in the north-western sub-structure of A2744. The filter stacks used are: Blue -- F090W+F115W+F150W, Green -- F200W+F277W and Red -- F356W+F410M+F444W. In the bottom panel, we mark the newly identified and numbered multiple image systems in green and multiple image candidates in yellow. These are listed in Tab.~\ref{tab:multiple_images1}. Cluster galaxies left free to vary in our SL model (see Tab.~\ref{tab:Galaxy-halo_results}) are shown in red. The critical curves of our SL model (see section~\ref{sec:result}) for source redshifts $z_{\mathrm{s}}=1.6881$, the redshift of system~1 \citep{mahler18}, and $z_{\mathrm{s}}=10$ are shown in blue and purple, respectively.}
    \label{fig:NW-extension}
\end{figure*}

\begin{figure*}
    \centering
    \includegraphics[width=\textwidth, keepaspectratio=true]{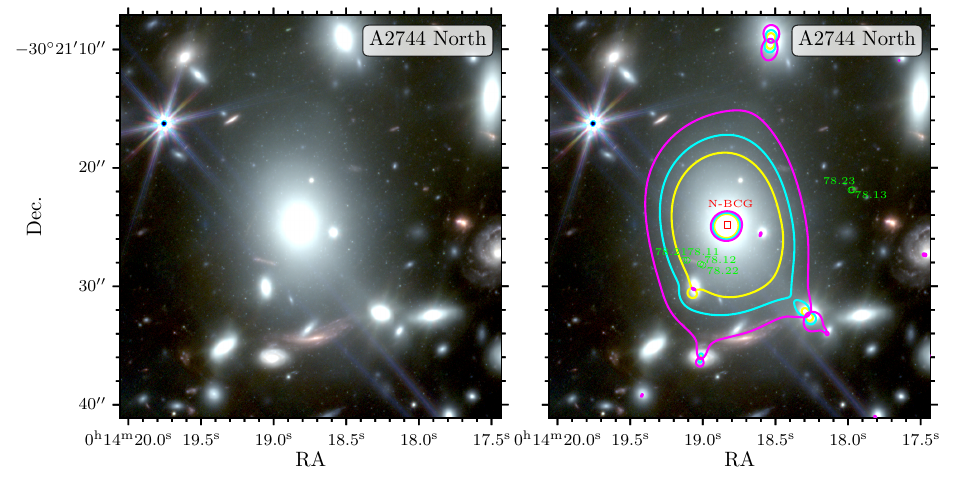}
    \caption{A $34\arcsec\times34\arcsec$ cutout of an UNCOVER color-composite image in the same filter configuration as in Fig.~\ref{fig:NW-extension}. In the right-hand panel, we mark the newly identified SL features around the northern sub-structure of A2744. The northern BCG, which is left free to vary in our SL model, is shown in red. We also show the critical curves of our SL model (see section~\ref{sec:result}) for source redshifts $z_{\mathrm{s}}=1$ in yellow, $z_{\mathrm{s}}=1.6881$ (see Fig.~\ref{fig:NW-extension}) in blue and $z_{\mathrm{s}}=10$ in purple. The critical curves from the current model do not pass where we would expect for system 78, suggesting that the system either lies at a lower redshift ($z_{\mathrm{s}}<1$), that it consists only of two images, or most probably, that the mass of this sub-halo is somewhat overestimated (see discussion in section~\ref{sec:result}). A red spiral galaxy south of the N-BCG is notably lensed, i.e. magnified and distorted, but is likely not multiply imaged.}
    \label{fig:N-extension}
\end{figure*}

\section{Observations and data} \label{sec:data}
As one of the six HFF clusters, Abell~2744 (A2744 hereafter) is a well-studied gravitational lensing system and has been targeted by many imaging and spectroscopic campaigns spanning multiple wavelength regimes. Among the ancillary data, the most relevant for our current study are of course the HFF HST images, which were recently complemented with the \textit{Beyond Ultra-deep Frontier Fields And Legacy Observations} \citep[BUFFALO; PIs: C. Steinhardt \& M. Jauzac;][]{steinhardt20} to extend the area of the HST coverage around the cluster center. In addition, there were extensive spectroscopic campaigns of the HFF clusters, both with the HST from space in e.g. the \textit{Grism Lens-Amplified Survey from Space} survey \citep[GLASS;][]{treu15} and from the ground with \textit{Keck} and ESO's \textit{Very Large Telescope} (VLT). The former was recently complemented with the first JWST spectroscopy of A2744 in the framework of the GLASS-JWST Early Release Science (ERS) program \citep{treu22} targeted at the cluster core. A2744 is also part of the \textit{ALMA Frontier Fields} survey \citep{gonzales-lopez17} which obtained 1.1\,mm imaging of the cluster core with the \textit{Atacama Large Millimeter/sub-millimeter Array} (ALMA) Band~6 and which will be complemented with additional observations covering the entire UNCOVER footprint (Program ID:~2022.1.00073.S; PI: S. Fujimoto).

In the following, we concentrate on the data sets that are directly used in this study to construct our SL model of the cluster, i.e. the UNCOVER JWST imaging in section~\ref{sec:uncover} and existing HST imaging and ground-based spectroscopy in section~\ref{sec:HFF}.

\subsection{JWST UNCOVER imaging} \label{sec:uncover}
The UNCOVER imaging of A2744 comprises extremely deep observations of the cluster and its surrounding field with the \textit{Near Infrared Camera} \citep[NIRCam;][]{rieke05,rieke23} aboard the JWST, achieving nominal $5\sigma$ depths of the order of $29.8$\,magnitudes before accounting for magnification. The magnification on average adds $\sim2$\,magnitudes to the depths in the SL regime, so that UNCOVER effectively reaches down to $\sim32$\,magnitude. The cluster was observed in six broad-band filters, F115W, F150W, F200W, F277W, F356W and F444W, and in one medium-band filter, F410M, for 4-6 hours each. The data were then reduced and drizzled into mosaics with the \texttt{grism redshift and line analysis software for space-based spectroscopy} (\texttt{grizli}\footnote{\url{https://github.com/gbrammer/grizli}}; Brammer et al. in prep.). These mosaics also incorporate the other public JWST imaging data available in the vicinity of A2744, namely the GLASS-JWST \citep{treu22} images obtained with NIRCam and the \textit{Near-Infrared Imager and Slitless Spectrograph} \citep[NIRISS;][Doyon et al. in prep.]{doyon12}, which in particular add F090W-band coverage in parts of the UNCOVER field, and the NIRCam imaging obtained in DDT program~2767 (PI: P. Kelly). For this work we use the \texttt{v5.4} UNCOVER mosaics, which are publicly available on the UNCOVER website\footnote{\url{https://jwst-uncover.github.io/DR1.html}} and will also soon be available on the \texttt{Barbara A. Mikulski Archive for Space Telescopes (MAST)}. We refer the reader to the UNCOVER survey paper \citep{bezanson22} for more details on the JWST data. With a field-of-view totaling $\sim45$\,arcmin$^{2}$, the UNCOVER mosaics represent the widest ultra-deep imaging coverage of A2744 to date. Note that deep \textit{Near Infrared Spectrograph} \citep[NIRSpec;][]{jakobsen22,ferruit22,boeker23} observations for UNCOVER are scheduled for July~2023.

In our analysis we furthermore use the first team-internal versions of the UNCOVER photometric catalogs, produced with a python implementation of \texttt{SExtractor} \citep[\texttt{SEP};][]{bertin96,barbary16,barabry18}, which will be published in subsequent UNCOVER data releases as detailed in \citet{weaver23}. These catalogs also contain photometric redshifts measured with \texttt{EAZY} \citep{brammer08} and gravitational magnifications derived from the lens model presented in this work.

\subsection{Archival HST and ground-based data} \label{sec:HFF}
In addition, the UNCOVER data products encompass HST mosaics drizzled to the same pixel-scale and WCS grid as the JWST imaging with \texttt{grizli}. Therefore, we also incorporate the HFF and extended-area BUFFALO observations in the \textit{Advanced Camera for Surveys} (ACS) F435W, F606W and F814W filters, and the \textit{Wide-Field Camera Three} (WCF3) F105W, F125W, F140W and F160W filters in our analysis.

We make use of the spectroscopic redshift catalogs constructed by \citet{mahler18}, refined in \citet{richard21}, and \citet{bergamini22}. These were measured using the deep and wide-field observations of A2744 obtained with the \textit{Multi-Unit Spectroscopic Explorer} \citep[MUSE;][]{bacon10} on the VLT in the framework of the MUSE guaranteed time observations (GTO; Program ID:~094.A-0115; PI: J. Richard). The data reduction is described in detail in \citet{richard21}. Additional redshifts for cluster galaxies used in this work were compiled in \citet{bergamini22} using data from GLASS, \textit{VIsible Multi-Object Spectrograph} \citep[VIMOS;][]{braglia09}, and \textit{Anglo-Australian Telescope} \citep[AAT;][]{owers11}.

\section{Strong lensing mass modeling} \label{sec:modeling}
In this section, we present our methods and the constraints used to construct the mass model of A2744. The lens modeling code is explained in section~\ref{sec:SL-code}. Our SL mass model of A2744 essentially takes two input components from the imaging and spectroscopy data: the cluster member galaxies (section~\ref{sec:cluster_members}) and the lensed multiple image systems (section~\ref{sec:multiple_images}). Finally, we explain our initial setup of the A2744 SL model in section~\ref{sec:SL_setup}.

\subsection{Modeling method} \label{sec:SL-code}
To model A2744 we use a code recently developed for fast parametric lens modeling of galaxy clusters. The method is an updated version of the previous (grid-based) parametric method presented in \citet{zitrin13a,zitrin13b,zitrin15a,zitrin21}, revised to now be completely ``analytic'' in the sense that the model is not limited by an input grid resolution. The code was tested on self-simulated mock mass distributions: Using the same mass parametrizations as used for the modeling, a suite of order $\sim10$ arbitrary mass distributions was generated by specifying for each one the input parameter values (namely core and truncation radii, and velocity dispersion of a reference galaxy, and the various DM halo parameters). We then planted sources behind the cluster and lensed them through the simulated mock cluster lens to predict the locations of their multiple images. SL models for these mock clusters were then constructed using these multiple images as input for the minimization, in order to test how well the input parameter values can be recovered. The code was found to work very well, recovering the input parameters to well below the 1$\sigma$ errors, both using lens and source plane minimizations. First successful models using this method with JWST data were published in \citet{pascale22} for SMACS~J0723.3-7327, where initial details about the modeling method were given, in \citet{hsiao22} and \citet{meena22b} for MACS~J0647.7+7015, and in \citet{williams22} for RX~J2129.4+0009. An HST-based A2744 model with this method was recently also used in \citet{roberts-borsani22}. The code is sometimes referred to as ``\texttt{Zitrin-Analytic}'' in these works.

The method is similar in essence to other, commonly used parametric techniques (e.g. \texttt{lenstool} \citep{kneib96,jullo07,jullo09}, \texttt{glafic} \citep{oguri10}, \texttt{GLEE} \citep{halkola06,grillo15}, etc.), and consists of two main mass components. The first mass component comprises the cluster galaxies, each galaxy modeled as a double Pseudo Isothermal Ellipsoid \citep[dPIE;][]{eliasdottir07} with projected surface mass density:

\begin{equation} \label{eq:dPIE}
    \Sigma(\xi)=\frac{\sigma_v^2r_{\mathrm{cut}}^2}{2G(r_{\mathrm{cut}}^2+r_{\mathrm{core}}^2)}\left[\left(1+\frac{r_{\mathrm{core}}^2}{\xi^2}\right)^{-\frac{1}{2}}-\left(1+\frac{r_{\mathrm{cut}}^2}{\xi^2}\right)^{-\frac{1}{2}}\right],
\end{equation}

where $\sigma_{v}$, $r_{\mathrm{core}}$ and $r_{\mathrm{cut}}$ are the velocity dispersion, core radius, and truncation radius, respectively, and $\xi$ is an elliptical coordinate defined as:

\begin{equation} \label{eq:xi}
    \xi^2=\frac{x^2}{(1-\epsilon)^2}+\frac{y^2}{(1+\epsilon)^2},
\end{equation}

with ellipticity $\epsilon$. In order to optimize the computation speed, we model most cluster galaxies as spherical ($\epsilon=0$). The ellipticity is used however for brightest cluster galaxies (BCGs) and eq.~\ref{eq:dPIE} is in practice implemented as in e.g. \citet{keeton01a} and \citet{oguri10}, where it is referred to as a ``Pseudo-Jaffe profile'' after \citet{jaffe83}. The total mass associated with a dPIE profile is then

\begin{equation} \label{eq:dPIE-mass}
    M_{\mathrm{tot}}=\frac{\pi\sigma_v^2}{G}\frac{r_{\mathrm{cut}}^2}{r_{\mathrm{cut}}+r_{\mathrm{core}}}.
\end{equation}

In order to drastically reduce the otherwise large number of free parameters that would arise from modeling each cluster galaxy individually, we use common scaling relations:

\begin{align}
    &\sigma_{v}=\sigma_{v,\star}\left(\frac{L}{L_{\star}}\right)^{\lambda} \label{eq:sigma-relation}, \\
    &r_{\mathrm{core}}=r_{\mathrm{core},\star}\left(\frac{L}{L_{\star}}\right)^{\beta} \label{eq:rcore-relation}, \\
    &r_{\mathrm{cut}}=r_{\mathrm{cut},\star}\left(\frac{L}{L_{\star}}\right)^{\alpha} \label{eq:rcut-relation},
\end{align}

which are based on e.g. the the Tully-Fisher \citep{tully77} and Faber-Jackson relations \citep{faber-jackson76} and the fundamental plane to tie the individual dPIE parameters of each galaxy to its luminosity $L$ and scale them with respect to a reference galaxy \citep[e.g.][]{halkola06,halkola07,jullo07,monna14,monna15}. The three parameters $\sigma_{v,\star}$, $r_{\mathrm{core},\star}$ and $r_{\mathrm{cut},\star}$ are those of a reference galaxy of luminosity $L_{\star}$ and can be left as free parameters in the model. Note that $r_{\mathrm{core},\star}$ is typically fixed however and only $\sigma_{\star}$ and $r_{\mathrm{cut},\star}$ are iterated for. The scaling parameters $\lambda$, $\beta$ and $\alpha$ can also be left as free parameters in the modeling. For example, \citet{jullo07} use $\lambda=0.25$, $\beta=0.5$ and $\alpha=0.5$, for a constant mass-to-light ratio, or $\alpha=0.8$ to more closely match the fundamental plane. As another example, \citet{monna15} for instance, do not adopt a core for the galaxies, and have $\lambda=0.3$ and $\alpha=0.4$. Note that unless these are specifically left to vary freely, the scaling relations \eqref{eq:sigma-relation}-\eqref{eq:rcut-relation} are also applied to the BCGs.

The second mass component consists of the diffuse cluster dark matter halos. These are modeled as Pseudo Isothermal Elliptical Mass Distributions \citep[PIEMDs; e.g.][]{jaffe83,keeton01a}

\begin{equation} \label{eq:PIEMD}
    \Sigma(\xi)=\frac{\sigma_v^2}{2G}\left(r_{\mathrm{core}}^2+\xi^2\right)^{-\frac{1}{2}}.
\end{equation}

One such diffuse cluster halo is usually sufficient for a relaxed cluster with a single prominent BCG, but several cluster halos are often needed to model merging or complex clusters, i.e. those containing various sub-structures and BCGs such as A2744 \citep[e.g.][]{jauzac15,mahler18,bergamini22}.

In addition, a two-component external shear can be applied, which can sometimes help account for external contributions not taken into account in the mass modeling. Source redshifts can be left as free parameters, especially for galaxies lacking spectroscopic redshifts, and the weight of individual cluster galaxies can also be left to be freely optimized. The latter can be applied in two ways -- either by incorporating the relative weight factor directly to the luminosity of the galaxy, affecting correspondingly all quantities derived from the scaling relations (this is the default choice), or if desired, only to the velocity dispersion of the galaxy, without affecting its other parameters. Similarly, the ellipticity parameters and core radii of the BCGs can also be left to be freely optimized. Finally, the code also allows for the incorporation of time delays, magnification ratios, and image parity information.

\subsection{Cluster member galaxies} \label{sec:cluster_members}

\begin{figure}
    \centering
    \includegraphics[width=\columnwidth, keepaspectratio=true]{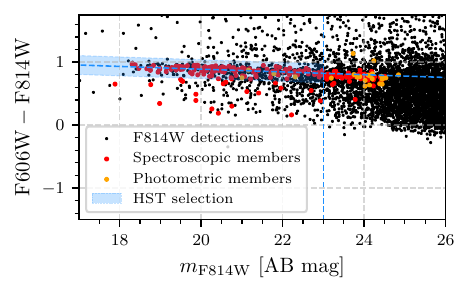}
    \caption{HST color-magnitude diagram of objects detected in A2744, showing the cluster's red sequence. The known cluster members from \citet{bergamini22} are shown in red and orange dots and our red sequence selection is shown as the blue shaded area.
    }
    \label{fig:red-sequence}
\end{figure}

We compile the list of cluster member galaxies hierarchically using all of the available data sets in the following way: We first include the 225 cluster members used in \citet{bergamini22}, 202 of which are spectroscopically confirmed to lie at the cluster redshift, $z_{\mathrm{d}}=0.308$. These were mostly derived from MUSE observations, as well as other ancillary catalogs listed in section~\ref{sec:HFF}. This spectroscopic sample is then used to calibrate red-sequence selection criteria aimed at identifying cluster member galaxies that lie beyond the MUSE coverage.

At the cluster redshift, the 4000\,\AA-break falls between the HST/ACS F606W and F814W bands. Because of that, these two filters are optimal for identifying the red sequence of A2744. We therefore perform a cluster member selection using the HST data (see section~\ref{sec:HFF}) which are much more extended than the initial HFF footprint owing to the wide-field BUFFALO and parallel field data. Galaxies are detected with \texttt{SExtractor} \citep{bertin96} in the F814W-band and photometry is measured in both ACS bands. The thus obtained color-magnitude diagram clearly shows the cluster's red sequence, as can be seen in Fig.~\ref{fig:red-sequence}. We fit a linear relation to the F606W-F814W color of the \citet{bergamini22} spectroscopic cluster members and use the result to select HST-detected galaxies in a color window of width 0.3\,magnitudes around that relation with magnitudes $m_{\mathrm{F814W}}\leq23$\,mag as cluster members. Note that this magnitude cut is only applied to the color-selected galaxies, we include all spectroscopically confirmed cluster members regardless of their brightness.

With this selection, we end up with a cluster member catalog containing 421 galaxies. Most notable are of course the two well-known brightest cluster galaxies (BCGs), BCG1 and BCG2, in the cluster core. In addition, the cluster has two BCGs at the position of the north-western extension, which we will denote NW-BCG1 and NW-BCG2, and one in the northern sub-structure which we will denote N-BCG. While these sub-structures were included as fixed or free DM clumps in previous SL models of A2744 \citep{mahler18,bergamini22}, we are able to confirm them as SL structures for the first time thanks to the UNCOVER imaging and constrain their parameters as will be shown in more detail in sections~\ref{sec:multiple_images} and~\ref{sec:result}. The wide HST coverage of the UNCOVER field reveals cluster member galaxies out to $4.11\arcmin$ from BCG1, which corresponds to 1.12\,Mpc at the cluster redshift. Together with the JWST UNCOVER data revealing a trove of SL constraints (see section~\ref{sec:multiple_images}), these now confirm the extended merging nature of A2744 already indicated by the previous X-ray and WL analyses \citep{merten11,medezinski16,jauzac16b}. Note that we will further discuss the large-scale structure of A2744 in light of our SL analysis results in section~\ref{sec:result}. The full catalog of cluster galaxies is included in the public release of our SL model of A2744 (see section~\ref{sec:result}).

\subsection{Multiple image systems} \label{sec:multiple_images}

\begin{table*}
    \caption{New multiple images and candidates identified in the northern and north-western extensions of A2744.}
    \begin{tabular}{lccccr}
    \hline
    ID      &   RA              &   Dec.            &   $z_{\mathrm{phot}}$~[97.5\,\%-range]  &   $z_{\mathrm{model}}$~[95\,\%-range] &   Remarks\\\hline
    65.11	&	00:14:13.570	&	-30:22:32.649	&	-	                	              &		2.798~$[2.473,3.106]$             &	Close to north-western BCG1.\\
	65.12	&	00:14:13.521	&	-30:22:36.422	&	-	                	              &		"                                 &	"\\
	c65.13	&	00:14:13.105	&	-30:22:45.219	&	3.483~$[3.258,3.741]$	              &		"                                 &	Candidate counter image.\\
	c65.13	&	00:14:12.979	&	-30:22:47.563	&	-	                	              &		"                                 &	"\\
	65.21	&	00:14:13.571	&	-30:22:32.916	&	-	                	              &		"                                 &	Close to north-western BCG1.\\
	65.22	&	00:14:13.530	&	-30:22:36.165	&	-	                	              &		"                                 &	"\\
	c65.23	&	00:14:13.084	&	-30:22:45.594	&	-	                	              &		"                                 &	Candidate counter image.\\
	c65.23	&	00:14:12.953	&	-30:22:47.671	&	-	                	              &		"                                 &	"\\
	65.31	&	00:14:13.563	&	-30:22:34.178	&	-	                	              &		"                                 &	Close to north-western BCG1.\\
	65.32	&	00:14:13.556	&	-30:22:35.004	&	-	                	              &		"                                 &	"\\\hline
	66.1	&	00:14:13.113	&	-30:22:25.817	&	2.227~$[1.994,2.421]$	              &		2.413~$[2.406,2.420]$             &	\\
	66.2	&	00:14:12.764	&	-30:22:33.592	&	-	                	              &		"                                 &	Close to north-western BCG1.\\
	66.3	&	00:14:12.556	&	-30:22:43.000	&	-	                	              &		"                                 &	\\\hline
	67.1	&	00:14:10.089	&	-30:22:28.583	&	2.423~$[2.288,2.510]$	              &		2.617~$[2.613,2.620]$             &	Bright system, invisible in HST data,\\
	67.2	&	00:14:10.279	&	-30:22:24.600	&	2.536~$[2.491,2.695]$	              &		"                                 &	\\
	67.3	&	00:14:10.632	&	-30:22:05.019	&	2.443~$[2.347,2.600]$	              &		"                                 &	\\\hline
	68.1	&	00:14:09.707	&	-30:22:30.242	&	2.524~$[2.495,2.698]$	              &		2.327~$[2.321,2.332]$             &	Bright system, invisible in HST data.\\
	68.2	&	00:14:09.681	&	-30:22:20.969	&	-	                	              &		"                                 &	Close to north-western BCG2.\\
	68.3	&	00:14:10.334	&	-30:22:05.143	&	2.443~$[2.234,2.621]$	              &		"                                 &	\\
	68.4	&	00:14:10.100	&	-30:22:21.444	&	-	                	              &		"                                 &	Radial image; close to north-western BCG2.\\\hline
	69.1	&	00:14:09.773	&	-30:22:33.161	&	2.423~$[2.228,2.630]$	              &		2.432~$[2.428,2.436]$             &	Bright system, invisible in HST data.\\
	69.2	&	00:14:09.823	&	-30:22:19.990	&	-	                	              &		"                                 &	Close to north-western BCG2.\\
	69.3	&	00:14:10.344	&	-30:22:08.778	&	2.325~$[2.224,2.351]$	              &		"                                 &	\\
	69.4	&	00:14:10.233	&	-30:22:20.856	&	-	                	              &		"                                 &	Radial image; close to north-western BCG2.\\\hline
	70.1	&	00:14:12.579	&	-30:22:26.363	&	-	                	              &		2.716~$[2.500,3.645]$             &	\\
	70.2	&	00:14:12.565	&	-30:22:26.675	&	-	                	              &		"                                 &	\\
	c70.3	&	00:14:12.109	&	-30:22:42.648	&	-	                	              &		"                                 &	\\\hline
	71.11	&	00:14:13.764	&	-30:22:41.450	&	-	                	              &		1.379~$[1.245,1.533]$             &	\\
	71.12	&	00:14:13.788	&	-30:22:40.884	&	-	                	              &		"                                 &	\\
	71.13	&	00:14:13.819	&	-30:22:39.388	&	-	                	              &		"                                 &	\\
	71.21	&	00:14:13.759	&	-30:22:41.574	&	-	                	              &		"                                 &	\\
	71.22	&	00:14:13.797	&	-30:22:40.744	&	-	                	              &		"                                 &	\\
	71.23	&	00:14:13.819	&	-30:22:39.597	&	-	                	              &		"                                 &	\\\hline
	72.1	&	00:14:11.915	&	-30:22:42.461	&	3.657~$[3.131,4.088]$	              &		3.544~$[3.531,3.565]$             &	\\
	72.2	&	00:14:12.268	&	-30:22:29.121	&	-	                	              &		"                                 &	\\
	72.3	&	00:14:12.552	&	-30:22:20.067	&	3.664~$[3.234,4.086]$	              &		"                                 &	\\\hline
	c73.1	&	00:14:11.508	&	-30:22:11.886	&	0.225~$[0.042,6.799]$	              &		-                                 &	\\
	c73.2	&	00:14:11.252	&	-30:22:23.351	&	4.037~$[1.173,5.340]$	              &		-                                 &	\\
	c73.3	&	00:14:10.810	&	-30:22:37.163	&	-	                	              &		-                                 &	\\\hline
	74.1	&	00:14:10.833	&	-30:22:29.697	&	2.707~$[1.324,4.563]$	              &		3.033~$[2.832,3.255]$             &	\\
	74.2	&	00:14:10.861	&	-30:22:28.911	&	4.451~$[1.913,4.921]$	              &		"                                 &	\\
	74.3	&	00:14:11.289	&	-30:22:10.152	&	2.228~$[1.834,2.523]$	              &		"                                 &	\\\hline
	c75.1	&	00:14:11.525	&	-30:22:10.488	&	2.040~$[0.409,11.036]$	              &		-                                 &	\\
	c75.2	&	00:14:11.174	&	-30:22:25.668	&	5.062~$[1.103,16.824]$	              &		-                                 &	\\
	c75.3	&	00:14:10.850	&	-30:22:35.548	&	-	                	              &		-                                 &	\\\hline
	c76.1	&	00:14:10.746	&	-30:22:34.898	&	-	                	              &		-                                 &	\\
	c76.2	&	00:14:11.046	&	-30:22:25.827	&	0.798~$[0.113,1.915]$	              &		-                                 &	\\
	c76.3	&	00:14:11.396	&	-30:22:08.988	&	-	                	              &		-                                 &	\\\hline
    \end{tabular}
    \par\smallskip
    \begin{justify}
        \texttt{Note.} -- \textit{Column~1}: Image ID. -- \textit{Columns~2 and~3}: Right ascension and declination (J2000.0). -- \textit{Column~4}: Photometric redshift and its 97.5\,\% range taken from the UNCOVER catalog (see section~\ref{sec:uncover}). -- \textit{Column~5}: Median redshift from our SL-model (see section~\ref{sec:result}). -- \textit{Column~6}: Additional remarks. Candidate images are marked with ``\emph{c}'' and are not used in the modeling.
    \end{justify}
    \label{tab:multiple_images1}
\end{table*}

\begin{table*}
    \contcaption{New multiple images and candidates identified in the northern and north-western extensions of A2744. We also include here two systems lensed by the main cluster core: the $z\simeq9.8$ system from \citep{zitrin14,roberts-borsani22}, and one dropout $z\sim7-8$ system \citep[i.e. system 53][]{atek14a,mahler18}, that we further confirm with the UNCOVER data. Note that system~53 is not included in the lens modeling at this stage.}
    \begin{tabular}{lccccr}
    \hline
    ID      &   RA              &   Dec.            &   $z_{\mathrm{phot}}$~[97.5\,\%-range]  &   $z_{\mathrm{model}}$~[95\,\%-range] &   Remarks\\\hline
	77.11	&	00:14:12.439	&	-30:22:39.334	&	-	                	              &		1.505~$[1.441,1.582]$             &	\\
	77.12	&	00:14:12.535	&	-30:22:35.316	&	-	                	              &		"                                 &	\\
	77.13	&	00:14:12.954	&	-30:22:22.743	&	-	                	              &		"                                 &	\\
	c77.21	&	00:14:12.493	&	-30:22:37.781	&	-	                	              &		"                                 &	\\
	c77.22	&	00:14:12.516	&	-30:22:36.783	&	-	                	              &		"                                 &	\\
	c77.31	&	00:14:12.501	&	-30:22:37.430	&	-	                	              &		"                                 &	\\
	c77.32	&	00:14:12.511	&	-30:22:37.016	&	-	                	              &		"                                 &	\\\hline
	78.11	&	00:14:19.108	&	-30:21:27.791	&	-	                	              &		2.571~$[2.049,2.767]$             &	Close to northern BCG.\\
	78.12	&	00:14:19.017	&	-30:21:28.108	&	-	                	              &		"                                 &	"\\
	78.13	&	00:14:17.971	&	-30:21:21.851	&	-	                	              &		"                                 &	\\
	78.21	&	00:14:19.130	&	-30:21:27.791	&	-	                	              &		"                                 &	Close to northern BCG.\\
	78.22	&	00:14:19.000	&	-30:21:28.209	&	-	                	              &		"                                 &	System redshift poorly constrained.\\
	78.23	&	00:14:17.980	&	-30:21:21.877	&	-	                	              &		"                                 &	\\\hline
	c79.1	&	00:14:09.818	&	-30:22:27.889	&	2.606~$[2.223,2.963]$	              &		-                                 &	\\
	c79.2	&	00:14:09.772	&	-30:22:22.089	&	-	                	              &		-                                 &	Close to north-western BCG2.\\
	c79.3	&	00:14:10.429	&	-30:22:03.847	&	-	                	              &		-                                 &	\\\hline
	c80.11	&	00:14:11.153	&	-30:22:35.005	&	-	                	              &		-                                 &	Candidate high-redshift object; possible F814W-dropout.\\
	c80.12	&	00:14:11.213	&	-30:22:32.917	&	-	                	              &		-                                 &	"\\
	c80.13	&	00:14:11.286	&	-30:22:31.056	&	-	                	              &		-                                 &	"\\
	c80.14	&	00:14:11.793	&	-30:22:10.935	&	0.706~$[0.541,13.853]$	              &		-                                 &	"\\
	c80.21	&	00:14:11.133	&	-30:22:35.298	&	11.839~$[0.993,15.925]$	              &		-                                 &	"\\
	c80.22	&	00:14:11.195	&	-30:22:33.287	&	-	                	              &		-                                 &	"\\
	c80.23	&	00:14:11.305	&	-30:22:30.000	&	15.104~$[1.056,19.585]$	              &		-                                 &	"\\
	c80.24	&	00:14:11.776	&	-30:22:11.062	&	12.357~$[0.537,14.560]$	              &		-                                 &	"\\\hline
	c81.1	&	00:14:10.713	&	-30:22:35.817	&	-	                	              &		-                                 &	\\
	c81.2	&	00:14:11.104	&	-30:22:23.761	&	-	                	              &		-                                 &	\\
	c81.3	&	00:14:11.367	&	-30:22:10.528	&	-	                	              &		-                                 &	\\\hline
    53.1    &   00:14:19.161    &   -30:24:05.664   &   7.188~$[7.142,7.165]$                 &     -                                 & Detected in HFF \citep{atek14a,mahler18}.\\
    53.2    &   00:14:20.051    &   -30:23:48.058   &   0.193~$[0.189,0.196]$                 &     -                                 & "; heavily affected by the ICL.\\
    53.3    &   00:14:23.331    &   -30:23:39.639   &   7.188~$[7.131,8.578]$                 &     -                                 & Newly identified in the UNCOVER imaging \citep{furtak22b}.\\\hline
    JD1A    &   00:14:22.202    &   -30:24:05.364   &   $z_{\mathrm{spec}}=9.76$              &     -                                 & see \citet{zitrin14,roberts-borsani22}\\
    JD1B    &   00:14:22.805    &   -30:24:02.700   &   "                                     &     -                                 & "\\
    cJD1C   &   00:14:18.607    &   -30:24:31.356   &   "                                     &     -                                 & \\\hline
    \end{tabular}
    \par\smallskip
    \begin{justify}
        \texttt{Note.} -- \textit{Column~1}: Image ID. -- \textit{Columns~2 and~3}: Right ascension and declination (J2000.0). -- \textit{Column~4}: Photometric redshift and its 97.5\,\% range taken from the UNCOVER catalog (see section~\ref{sec:uncover}). -- \textit{Column~5}: Median redshift from our SL-model (see section~\ref{sec:result}). -- \textit{Column~6}: Additional remarks. Candidate images are marked with ``\emph{c}'' and are not used in the modeling.
    \end{justify}
    \label{tab:multiple_images2}
\end{table*}

The SL features at the center of A2744 have been extensively studied with the exceptionally deep HFF imaging and GLASS and MUSE spectroscopy (see section~\ref{sec:HFF}). These data yielded a total of 60 multiply imaged galaxies in A2744, producing 188 images in the cluster core \citep[][]{mahler18,richard21}. This list was then recently refined to a secure, purely spectroscopic, sample by \citet{bergamini22}. For our SL model, we use this secure spectroscopic sample of 90 multiple images belonging to 30 sources to constrain the main cluster core and refer the reader to \citet{bergamini22} (Tab.~A.1) for a complete list of images and redshifts. To this we add the triply-imaged \citet{zitrin14} $z\sim10$ object which was recently confirmed with JWST/NIRSpec spectroscopy at $z_{\mathrm{spec}}\simeq9.76$ \citep[][]{roberts-borsani22}. We also identify a notable triply imaged dropout system at $z\sim7-8$. Two of its images were previously reported in HFF high-redshift samples \citep{atek14a,atek15a} and identified as a photometric multiple image system in \citet{mahler18}, system 53. The new UNCOVER data now not only reveal a third image to this system but together with the SL model also help us further to confirm its high-redshift nature. Furthermore, in the JWST observations the object appears to be extremely red and compact with a point-like morphology, suggesting it is potentially an obscured Active Galactic Nucleus \citep[AGN;][]{furtak22b}. Note that this new image was identified after constructing the current lens model and will therefore only be included in future versions released in future UNCOVER data releases.

The new ultra-deep UNCOVER imaging of A2744 reveals spectacular new strongly lensed multiple image systems in the northern and north-western extensions of the cluster. While these extended sub-structures are known from X-ray and WL analyses \citep{merten11,medezinski16,jauzac16b}, they have so far not yet been confirmed to be dense enough to produce clear SL multiple images. Using the UNCOVER mosaics (see section~\ref{sec:uncover}), we identify 17 new multiple image systems and candidate systems in the north-western and one in the northern sub-structure of the cluster in a by-eye search considering their color, visual aspect, relative geometric position, and parity. The new systems are listed and summarized in Tab.~\ref{tab:multiple_images1} and can be seen in Figs.~\ref{fig:NW-extension} and~\ref{fig:N-extension} for the north-western and northern sub-structures. For systems that lack clear point-like features, we identify locations which show other unique features repeating in all counter images, such as e.g. places were the arc gets thicker or thinner, and use these as approximate image positions. The northern sub-cluster has only one clearly identified multiple image system, system 78. Note that we continue from the numbering scheme of \citet{mahler18}, which was also perpetuated in \citet{bergamini22}, i.e. we start at 65. Some of these multiple image systems, in particular systems 65, 71 and 78, present several sub-clumps in the JWST imaging, each of which can be used as an independent source in the SL modeling. Taking these into account, we identify a total of 75 new multiple images and image candidates in the north-western and northern sub-structures of A2744. Note that we do not use candidate multiple images, marked ``c'' in Tab.~\ref{tab:multiple_images1}, in the lens model (see section~\ref{sec:SL_setup}).

Some of these new multiply imaged galaxies are of particular interest for other studies. For example, the four relatively bright red systems 66 to 69 and the fainter candidate system c79 (see Fig.~\ref{fig:NW-extension}) are completely invisible in the HST data but all have well-constrained photometric redshifts in the preliminary UNCOVER stellar population analysis (see section~\ref{sec:uncover}) $z_{\mathrm{phot}}\simeq2.2-2.5$. This hints as to their probably very dusty nature and their apparent spatial proximity could indicate an overdensity at $z\sim2.5$. Indeed, the JWST has already revealed a population of such red galaxies around the cosmic noon and even up to the epoch of cosmic reionization \citep[e.g.][]{nonino22,fudamoto22,nelson22,barrufet22,zavala22,naidu22c,glazebrook22}, similarly to these galaxies. Systems 68 and 69 each show a relatively bright radial image close to the north-western BCG2, as can also be seen in Fig.~\ref{fig:NW-extension}, which are of particular value to constrain the mass distribution of that cluster galaxy. We have furthermore tentatively identified a candidate multiply imaged high-redshift source, system c80, which shows various multiply imaged clumps. These images show flux in the UNCOVER bands, i.e. in F115W and red-ward, but are invisible in the HST filters which could mean that they are F814W-dropouts at $z\sim7-8$. This is also supported by our lens model (see the high-redshift critical curve in Fig.~\ref{fig:NW-extension}) presented in section~\ref{sec:result}. Note however that the BUFFALO coverage in this part of the cluster is much shallower than the central HFF observations such that a low-redshift solution cannot be ruled out at this stage. We will therefore need spectroscopic observations of this object with JWST/NIRSpec to accurately determine its redshift and perhaps multiply imaged nature.

\subsection{Model setup} \label{sec:SL_setup}
Since A2744 is known to be a complex merging cluster, we follow the previously published SL models \citep[e.g.][]{jauzac15,mahler18,bergamini22} and place two PIEMD DM halos in the cluster core, initially centered on the two BCGs but with a free center position that is nonetheless limited to the vicinity ($\lesssim3\arcsec$) of the BCGs. In addition, we leave the weights of the two BCGs free as well as those of several other galaxies situated close to multiple images (see Fig.~\ref{fig:crit_curves}). For BCG1 (see section~\ref{sec:cluster_members}) we also leave the ellipticity parameters free but fix them to the \texttt{SExtractor}-measured values for BCG2.

In order to model the newly discovered SL structures in the north-west and in the north (see sections~\ref{sec:cluster_members} and~\ref{sec:multiple_images}), we place two PIEMD halos in the north-western sub-structure, centered on the two BCGs, and another one fixed on the northern BCG. The centers of the two north-western halos are left free within $3\arcsec$ of their respective BCG. We also allow for the ellipticity parameters, weight and the core radius of the NW-BCG2 to vary since it lies very close to the radial images of systems 68 and 69. A cluster galaxy next to images 68.2 and 69.2 is also left with a free weight (see Fig.~\ref{fig:NW-extension}). Finally, we fix the value of the reference galaxy core radius in eq.~\eqref{eq:rcore-relation} to $r_{\mathrm{core},\star}=0.2$\,kpc, similar to other SL works \citep[e.g.][]{richard14}. Such values, namely $r_{\mathrm{core},\star}=0.2$\,kpc, are typical of elliptical galaxies \citep[e.g.][]{wallington93}.

Our SL model is constrained with a total of 135 multiple images, 92 of which have spectroscopic redshifts and are situated in the cluster core as detailed in section~\ref{sec:multiple_images}. Among the remaining images, 37 lie in the north-western extension and 6 in the northern sub-structure. For all systems with available spectroscopic redshifts, the redshift is fixed to the measured one in the model. In addition, as can be seen in Tab.~\ref{tab:multiple_images1}, systems 66 to 69 and system 72 have relatively well-constrained photometric redshifts. We therefore leave these redshifts as free parameters in the model but limit them to the 97.5\,\% ranges of their photometric redshifts, which are relatively narrow ($\Delta z\lesssim0.4$). All other source redshifts are left free to vary between $z_{\mathrm{s}}=0.8$ and $z_{\mathrm{s}}=10$. Note that no parity information is used in this model.

The model is optimized using a Monte-Carlo Markov Chain (MCMC) analysis in two steps: First by running $\sim150$ relatively short chains of several thousand steps each to explore the parameter space, derive the covariance matrix encoding the correlations between the different parameters, and potentially to approach the global minimum of the source plane $\chi^2$-function defined as

\begin{equation} \label{eq:SP_Chi2}
    \chi^2=\sum_{i}{\frac{(\Vec{\beta}_i-\Vec{\beta}_0)^2}{\mu_i^2\sigma_i^2}}+\sigma_pp,
\end{equation}

where $\Vec{\beta}_i$ represents the source position of each multiple image, $\Vec{\beta}_0$ the common modeled barycenter source position of the images belonging to the same system, and $\mu_i$ and $\sigma_i$ the magnification and positional uncertainty of each multiple image. The second term represents the forced parities where $p$ is the number of images modeled with the wrong parity and $\sigma_p$ a constant factor to add to the $\chi^2$ for each wrong parity. Note that we perform the minimization in the source plane which in principle should perform just as well as a lens plane minimization \citep{keeton10}, as is also evident by the quality of some HFF SL models that were built using a source plane minimization, but has the advantage of being significantly faster. The result from this multi-chain first step is then further refined with two or more long MCMC chains, each of several $\sim10^4$ steps and a decreasing ``temperature'', enabling in practice some annealing. We assume a positional uncertainty of $\sigma_i=0.5\arcsec$ for all multiple images used as constraints. We can also evaluate the $\chi^2$ in the lens plane, defined as

\begin{equation} \label{eq:LP_Chi2}
    \chi^2=\sum_{i}{\frac{(\Vec{\theta}_i-\Vec{\theta}_{i,\mathrm{model}})^2}{\sigma_i^2}}+\sigma_pp,
\end{equation}

where $\Vec{\theta}_i$ is the observed position of the $i$-th image and $\Vec{\theta}_{i,\mathrm{model}}$ its predicted position, which should be roughly similar to the source-plane $\chi^2$ \citep{keeton10}.

Note that although the nominal positional uncertainty adopted for the multiple images in the $\chi^2$ calculations in \eqref{eq:SP_Chi2} and \eqref{eq:LP_Chi2} is taken as 0.5\arcsec, the statistical uncertainties of the resulting model parameters are calculated from an MCMC chain that is run with effectively $\sim3.5\times$ that value, 1.76\arcsec. In practice, this is done by adopting a higher temperature when estimating the parameter uncertainties than the nominal one used in the minimization of the final chain, which we have found to be representative of the true parameter uncertainties. Note that this only affects the uncertainty estimate of the parameters and not the best-fit value or the $\chi^2$ minimization. This factor was estimated previously on mock clusters by repeating the minimization 50 times while perturbing all multiple image positions with a Gaussian of width $\sigma=0.5\arcsec$ in both coordinates in order to assess how the multiple images' positional uncertainty affects the parameter posterior distribution.

\section{Results and Discussion} \label{sec:result}

\begin{figure*}
    \centering
    \includegraphics[width=\textwidth, keepaspectratio=true]{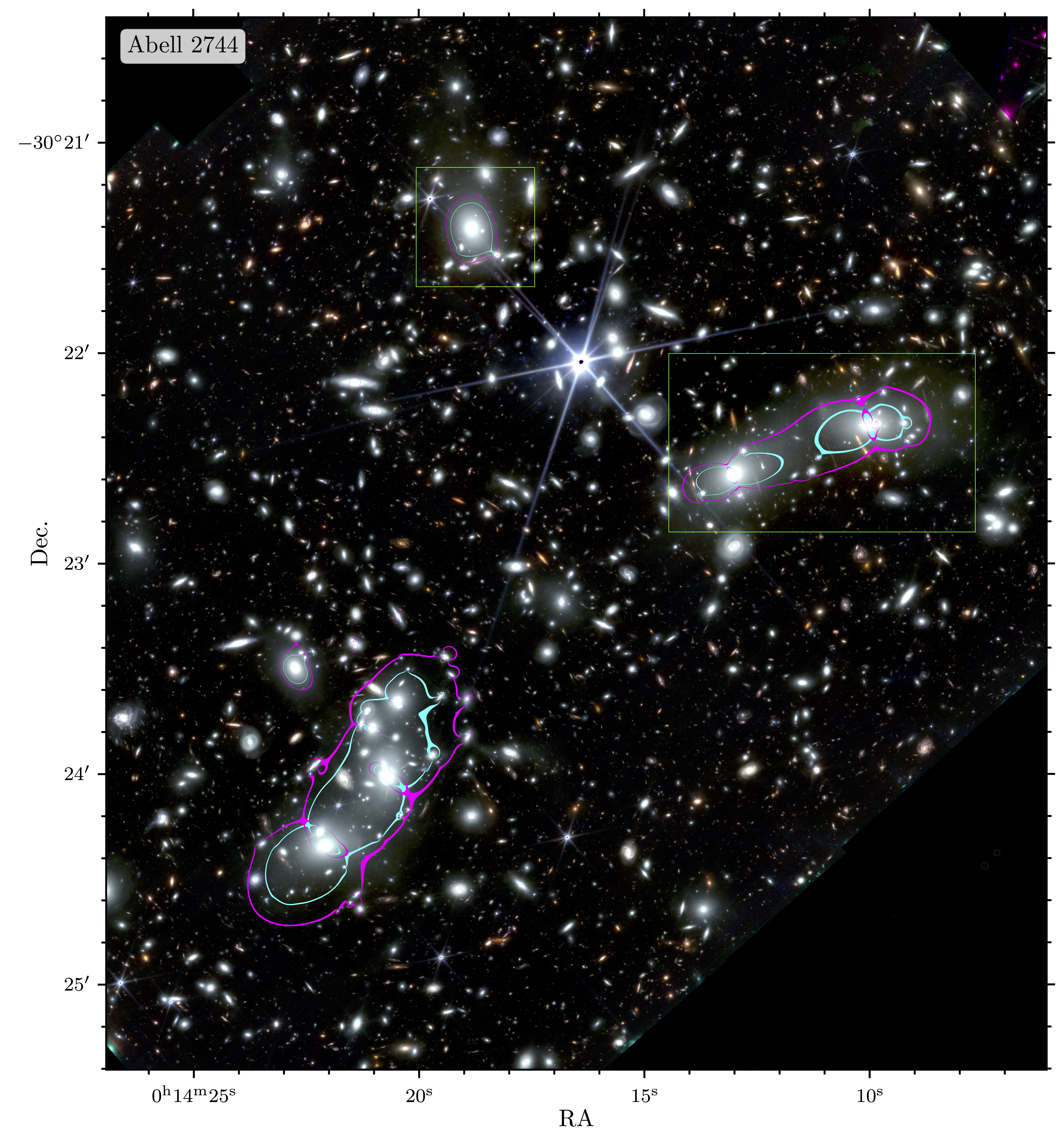}
    \caption{A $4.5\arcmin\times5.0\arcmin$ cutout of an UNCOVER/NIRCam composite-color image of A2744 in the same filter configuration as in Figs.~\ref{fig:NW-extension} and~\ref{fig:N-extension}. Cluster galaxies left free to vary in our SL model (see section~\ref{sec:SL_setup} and Tb.~\ref{tab:Galaxy-halo_results}) are shown in red. Overlaid we show the critical curves of our SL model for source redshifts $z_{\mathrm{s}}=1.6881$ (corresponding to system~1) and $z_{\mathrm{s}}=10$ in blue and purple respectively as in Fig~\ref{fig:NW-extension}. The area between the main cluster and the north-western sub-structure in particular has high magnifications of order $\mu\gtrsim4$ for sources at $z_{\mathrm{s}}=10$ (see Fig.~\ref{fig:magmap_z10}). The green rectangles outline the fields shown in more detail in Figs.~\ref{fig:NW-extension} and~\ref{fig:N-extension}. Note, a vectorized full UNCOVER resolution (0.04\arcsec/pix) version of this figure is available in the online supplemental material of this paper.}
    \label{fig:crit_curves}
\end{figure*}

\begin{figure}
    \centering
    \includegraphics[width=\columnwidth, keepaspectratio=true]{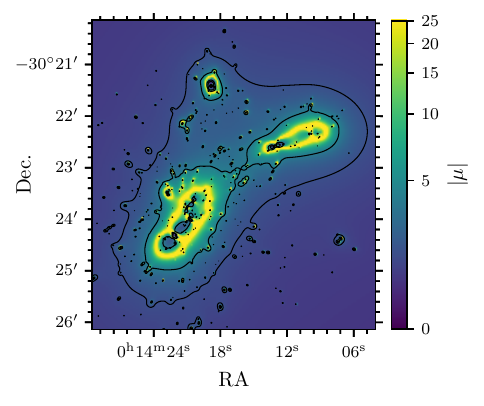}
    \caption{Magnification map of our final model for a source redshift $z_{\mathrm{s}}=10$. The black contours represent magnification thresholds of $\mu=2$ and $\mu=4$. The area between the main cluster and the extended sub-structures forms contiguous sheet of $\mu>2$ which goes up to $\mu\gtrsim4$ between the main cluster and the north-western extension.}
    \label{fig:magmap_z10}
\end{figure}

\begin{table*}
    \caption{Median DM halo PIEMD parameters, defined in eqs.~\eqref{eq:xi} and~\eqref{eq:PIEMD} in section~\ref{sec:SL-code}, and their $1\sigma$-errors for our SL model of A2744.}
    \begin{tabular}{lcccccc}
    \hline
    Halo            &   RA                              &   Dec.                             &   $\epsilon$  &   $\theta$ [Degrees]   &   $\sigma_v$ [$\frac{\mathrm{km}}{\mathrm{s}}$] &   $r_{\mathrm{core}}$ [kpc]\\\hline
                    &   \multicolumn{6}{c}{\textit{Main cluster core}}\\\hline
    Main-1          &   00:14:22.17$_{-0.01}^{+0.01}$   &   -30:24:18.32$_{-0.22}^{+0.31}$   &   $0.41_{-0.03}^{+0.02}$   &  $45_{-1}^{+4}$   &   $681_{-16}^{+34}$   &  $23_{-2}^{+4}$\\
    Main-2          &   00:14:20.80$_{-0.01}^{+0.03}$   &   -30:24:00.85$_{-0.64}^{+0.29}$   &   $0.45_{-0.02}^{+0.01}$   &  $73_{-2}^{+3}$   &   $807_{-72}^{+45}$   &  $67_{-20}^{+8}$\\\hline
                    &   \multicolumn{6}{c}{\textit{North-western extension}}\\\hline
    North-West-1    &   00:14:13.16$_{-0.01}^{+0.02}$   &   -30:22:35.65$_{-0.13}^{+0.34}$   &   $0.55_{-0.02}^{+0.02}$   &  $19_{-7}^{+3}$   &   $407_{-18}^{+21}$   &  $5_{-1}^{+1}$\\
    North-West-2    &   00:14:10.08$_{-0.01}^{+0.03}$   &   -30:22:19.90$_{-0.38}^{+0.22}$   &   $0.18_{-0.04}^{+0.02}$   &  $-8_{-5}^{+3}$   &   $906_{-73}^{+30}$   &  $62_{-16}^{+4}$\\\hline
                    &   \multicolumn{6}{c}{\textit{Northern extension}}\\\hline
    North           &   00:14:18.83                     &   -30:21:24.84                     &   $0.10_{-0.02}^{+0.02}$   &  $-73_{-5}^{+2}$  &   $486_{-16}^{+17}$    &  $2_{-1}^{+2}$\\\hline
    \end{tabular}
    \par\smallskip
    \begin{justify}
        \texttt{Note.} -- \textit{Column~1}: DM halo designation -- \textit{Columns~2 and~3}: Right ascension and declination. The northern halo is fixed onto it's BCG's position (see section~\ref{sec:SL_setup}). -- \textit{Column~4}: Ellipticity defined as $\epsilon=\frac{b-a}{a+b}$ where $a$ and $b$ are the semi-major and -minor axes respectively. -- \textit{Column~5}: Position angle, counter-clockwise with respect to the east-west axis. -- \textit{Column~6}: Velocity dispersion of the PIEMD profile. -- \textit{Column~7}: Core radius of the PIEMD profile.
    \end{justify}
    \label{tab:DM-halo_results}
\end{table*}

\begin{table*}
    \caption{Median dPIE parameters, defined in eqs.~\eqref{eq:dPIE} and~\eqref{eq:xi} in section~\ref{sec:SL-code}, and their $1\sigma$-errors of the BCGs and the galaxies left free in our SL model of A2744.}
    \begin{tabular}{lcccccccc}
    \hline
    Galaxy  &   RA$^{\mathrm{a}}$   &   Dec.$^{\mathrm{a}}$ &   $\sigma_v~[\frac{\mathrm{km}}{\mathrm{s}}]$ &   $r_{\mathrm{core}}$~[kpc]                       &    $r_{\mathrm{cut}}$~[kpc]$^{\mathrm{b}}$    &   $\epsilon$                  &   $\theta$~[Degrees]           &   $M_{\mathrm{tot}}~[10^{12}\,\mathrm{M}_{\odot}]$\\\hline
            &   \multicolumn{8}{c}{\textit{Main cluster core}}\\\hline
    BCG1    &   00:14:22.09         &   -30:24:20.73        &   $233_{-15}^{+30}$                  &   $0.525_{-0.016}^{+0.033~\mathrm{b}}$ &    $126\pm20$                        &  $0.14\pm0.02$       &   $31_{-3}^{+3}$     &   $4.98_{-1.67}^{+2.01}$\\
    BCG2    &   00:14:20.70         &   -30:24:00.59        &   $311_{-25}^{+26}$                  &   $0.499_{-0.015}^{+0.029~\mathrm{b}}$ &    $121\pm19$                        &  $0.09^{\mathrm{a}}$ &   $-67^{\mathrm{a}}$ &   $8.52_{-2.96}^{+3.00}$\\
    G1      &   00:14:22.75         &   -30:23:29.94        &   $341_{-19}^{+23}$                  &   $0.469_{-0.013}^{+0.026~\mathrm{b}}$ &    $115\pm18$                        &  -                   &   -                  &   $9.73_{-3.18}^{+3.27}$\\
    G2      &   00:14:20.49         &   -30:23:39.38        &   $225_{-22}^{+16}$                  &   $0.364_{-0.007}^{+0.014~\mathrm{b}}$ &    $94\pm14$                         &  -                   &   -                  &   $3.44_{-1.24}^{+1.16}$\\
    G3      &   00:14:21.03         &   -30:23:47.11        &   $200_{-16}^{+16}$                  &   $0.326_{-0.005}^{+0.010~\mathrm{b}}$ &    $86\pm13$                         &  -                   &   -                  &   $2.48_{-0.85}^{+0.86}$\\
    G4      &   00:14:22.20         &   -30:24:16.61        &   $161_{-16}^{+30}$                  &   $0.287_{-0.003}^{+0.007~\mathrm{b}}$ &    $77\pm12$                         &  -                   &   -                  &   $1.45_{-0.53}^{+0.70}$\\\hline
            &   \multicolumn{8}{c}{\textit{North-western extension}}\\\hline
    NW-BCG1 &   00:14:13.01         &   -30:22:34.72        &   $201_{-8}^{+9~\mathrm{b}}$       &   $0.562_{-0.019}^{+0.037~\mathrm{b}}$ &    $134\pm21$                        &  $0.07^{\mathrm{a}}$ &   $38^{\mathrm{a}}$  &   $3.92_{-1.25}^{+1.26}$\\
    NW-BCG2 &   00:14:10.01         &   -30:22:20.79        &   $218_{-23}^{+18}$                  &   $0.614_{-0.050}^{+0.073}$              &    $126\pm19$                        &  $0.16\pm0.01$       &   $-16_{-6}^{+2}$    &   $4.36_{-1.63}^{+1.52}$\\
    G5      &   00:14:09.81         &   -30:22:20.56        &   $108_{-7}^{+6}$                    &   $0.095_{-0.003}^{+0.007~\mathrm{b}}$ &    $31\pm5$                          &  -                   &   -                  &   $0.27_{-0.09}^{+0.09}$\\\hline
            &   \multicolumn{8}{c}{\textit{Northern extension}}\\\hline
    N-BCG   &   00:14:18.83         &   -30:21:24.83        &   $196_{-8}^{+9~\mathrm{b}}$       &   $0.533_{-0.017}^{+0.034~\mathrm{b}}$ &    $128\pm20$                        &  $0.14^{\mathrm{a}}$ &   $-70^{\mathrm{a}}$ &   $3.56_{-1.14}^{+1.14}$\\\hline
    \end{tabular}
    \par
    \begin{justify}
        $^{\mathrm{a}}$ Fixed to the \texttt{SExtractor}-measured values (see section~\ref{sec:cluster_members}).\\
        $^{\mathrm{b}}$ Scaled to the galaxy luminosity using the relations \eqref{eq:sigma-relation}-\eqref{eq:rcut-relation}.
        
        \par\smallskip\texttt{Note.} -- \textit{Column~1}: Galaxy designation. -- \textit{Columns~2 and~3}: Right ascension and declination. -- \textit{Column~4}: Velocity dispersion of the dPIE profile. -- \textit{Column~5}: Core radius of the dPIE profile. -- \textit{Column~6}: Truncation radius of the dPIE profile. -- \textit{Column~7}: Ellipticity defined as $\epsilon=\frac{b-a}{a+b}$ where $a$ and $b$ are the semi-major and -minor axes respectively. All non-BCG-type galaxies are assumed to be spherical (see section~\ref{sec:SL-code}). -- \textit{Column~8}: Position angle, counter-clockwise with respect to the east-west axis. -- \textit{Column~9}: Total mass of the dPIE profile computed from the model parameters following eq.~\ref{eq:dPIE-mass}.
    \end{justify}
    \label{tab:Galaxy-halo_results}
\end{table*}

Our resulting SL model of A2744 has a final \emph{lens plane $\chi^2$ \eqref{eq:LP_Chi2}} of $\chi^2=234$ and an average, lens-plane image reproduction RMS of $0.66\arcsec$. Our model has 161 SL constraints (i.e. $2\times135$, the number of images, minus $2\times46$, the number of systems used in the modeling, minus 15, the number of systems whose redshift was left free, after equation~1 in \citealt{Kneib1993}), and 61 free parameters, which results in 102 degrees of freedom. This nominally leads to a reduced lens plane $\chi^{2}$ of $\simeq2.3$, i.e. somewhat above but of order unity. For comparison, our image reproduction RMS is of the same order as quoted for the \citet{mahler18} model ($0.67\arcsec$) and $1.8\times$ higher than the value quoted for the \citet{bergamini22} model ($0.37\arcsec$). Note however that we include a multitude (15) of new multiple image systems without precise redshift measurements to constrain the outer sub-structures, as opposed to their secure purely spectroscopic sample in the cluster core. Indeed, e.g. \citet{johnson16} found that the RMS-scatter in the lens plane may degrade when less spectroscopic systems are used for the modeling. Moreover, caution should be taken when comparing RMS values, which may be calculated somewhat differently in different works or between different modelers. 

We show the critical curves for sources at redshifts $z_{\mathrm{s}}=1.6881$ and $z_{\mathrm{s}}=10$ in Fig.~\ref{fig:crit_curves}. The redshift $z_{\mathrm{s}}=1.6881$ corresponds to the redshift of system~1 \citep{mahler18} and represents the redshift to which the distance ratios are normalized in our model. More detailed zoomed-in versions on the extended sub-structures in the north-west and in the north can be seen in Figs.~\ref{fig:NW-extension} and~\ref{fig:N-extension}, respectively. The figure illustrates how the inclusion of the northern and north-western extensions increases the total critical area of A2744 by a factor $\sim1.5$. We indeed find a total critical area of $A_{\mathrm{crit}}=0.67\,\mathrm{arcmin}^{2}$ which translates into an effective Einstein radius of $\theta_E=27.7\arcsec\pm2.8\arcsec$ for a source at $z_{\mathrm{s}}=2$. The total cluster mass enclosed by those curves is $M(<\theta_E)=(1.07\pm0.15)\times10^{14}\,\mathrm{M}_{\odot}$. In particular the area between the main cluster and north-western extension forms a high-magnification `bridge' where the magnification remains $\mu\gtrsim4$ for sources at $z_{\mathrm{s}}=10$ as can be seen in Fig.~\ref{fig:magmap_z10}. Individually, the main cluster has a $z_{\mathrm{s}}=2$ critical area of $A_{\mathrm{crit,main}}=0.47\,\mathrm{arcmin}^2$ ($\theta_{E,\mathrm{main}}=23.2\arcsec\pm2.3\arcsec$) comprising a mass of $M(<\theta_{E,\mathrm{main}})=(7.7\pm1.1)\times10^{13}\,\mathrm{M}_{\odot}$. The northern and north-western substructures each have critical areas of $A_{\mathrm{crit,N}}=0.05\,\mathrm{arcmin}^2$ ($\theta_{E,\mathrm{N}}=7.4\arcsec\pm0.7\arcsec$) and $A_{\mathrm{crit,NW}}=0.15\,\mathrm{arcmin}^2$ ($\theta_{E,\mathrm{NW}}=13.1\arcsec\pm1.3\arcsec$), enclosing masses of $M(<\theta_{E,\mathrm{N}})=(0.8\pm0.1)\times10^{13}\,\mathrm{M}_{\odot}$ and $M(<\theta_{E,\mathrm{NW}})=(2.2\pm0.3)\times10^{13}\,\mathrm{M}_{\odot}$ respectively. In the above we adopt nominal errors of 10\% and 14\% for the effective Einstein radii and enclosed masses, respectively, which encompass the typical systematic scatter of these values between different models \citep[e.g.][]{zitrin15a}. The statistical errors that we obtain from our model are smaller, of the order $\sim2-3$\,\%\ for the effective Einstein radius and $\sim5$\,\%\ for the enclosed total mass.

The median PIEMD parameters of the five DM halos in our model are listed in Tab.~\ref{tab:DM-halo_results}. Note that the northern halo seems to be unexpectedly massive given the apparent dearth of multiple image systems in that region (see section~\ref{sec:multiple_images}). This particular sub-structure is therefore probably poorly constrained due to its lack of many multiple images and spectroscopic redshifts to anchor the model in that region. Because of that, we surmise that this halo in particular might be overestimated in our model. Our results in the main cluster agree well with \citet{mahler18} and \citet{bergamini22} which is unsurprising since we use a similar setup and rely on the same set of multiple images and spectroscopic redshifts to constrain this region of the cluster. In particular, the \citet{mahler18} model, which achieved a similar lens plane RMS as ours (see above), finds an effective Einstein radius of $\simeq23.9\arcsec$ for $z_{\mathrm{s}}=4$ \citep[given in][]{richard21} in the main cluster core where our model yields $\theta_{E,\mathrm{main}}=26.2\arcsec\pm2.6\arcsec$ for the same source redshift. These two models therefore concur within the uncertainties. We also refer the reader to Fig.~1 in \citet{roberts-borsani22} where we overlaid the $z_{\mathrm{s}}=10$ critical curves of our preliminary main cluster core model alongside those of the \citet{zitrin14} and \citet{bergamini22} models. Furthermore, we find a total cluster mass of $M(r<200\,\mathrm{kpc})=(1.60\pm0.22)\times10^{14}\,\mathrm{M}_{\odot}$ within 200\,kpc of BCG2 (see Fig.~\ref{fig:crit_curves}) which also agrees with the value found in \citet{bergamini22} within the uncertainties. The two models by \citet{mahler18} and \citet{bergamini22} also include the extended sub-structures in the north and the north-west based on the previous predictions from WL and the BUFFALO HST imaging,  but do not constrain them directly due to the lack on SL features in that region prior to the UNCOVER imaging. With the new multiple images identified in this study, however, we find much less massive DM halos than \citet{mahler18} and slightly less massive halos than \citet{bergamini22} in these structures. In particular, the northern DM halo has a much lower mass in our model (see Tab.~\ref{tab:DM-halo_results}) than in both \citet{mahler18} and \citet{bergamini22} and even so, judging by the single multiply imaged system found around that clump (see Fig.~\ref{fig:N-extension}), we suspect that the mass of this sub-structure is overestimated, as explained above.

The resulting parameters for the BCGs and cluster galaxies left free in our model (see section~\ref{sec:SL_setup}) are listed in Tab.~\ref{tab:Galaxy-halo_results}. We find the reference galaxy of (fixed) luminosity $M_{\star}=-21.05$ (in absolute magnitudes) to have a velocity dispersion $\sigma_{v,\star}=117\pm4\,\frac{\mathrm{km}}{\mathrm{s}}$ and a truncation radius $r_{\mathrm{cut},\star}=57\pm9\,\mathrm{kpc}$. The median scaling relations eqs.~\eqref{eq:sigma-relation}-\eqref{eq:rcut-relation} are $\lambda=0.40_{-0.01}^{+0.02}$, $\beta=0.77_{-0.02}^{+0.05}$ and $\alpha=0.63_{-0.02}^{+0.01}$. Note that these seem to work very well, as is evident by the fact that multiple image systems that lie very close to cluster galaxies, such that their lensing signal is dominated by them (e.g., systems 65 and 71; see Fig.~\ref{fig:NW-extension}), are exceptionally well reproduced by our model. In addition, the $\lambda$-parameter agrees perfectly with the value of $\simeq0.4$ measured in \citet{bergamini22} from the internal stellar kinematics of cluster galaxies \citep[][]{bergamini19,bergamini21}. While our result for the power-law index of the scaling relation for the truncation radius disagrees with the $\alpha\simeq0.41$ kinematic measurement in \citet{bergamini22}, we also find a much higher $r_{\mathrm{cut},\star}$ amplitude in our model despite our reference galaxy being fainter. Using eq.~\eqref{eq:dPIE-mass}, we compute the total mass of our reference galaxy $M_{\mathrm{tot},\star}=(5.7\pm1.8)\times10^{11}\,\mathrm{M}_{\odot}$ which, together with its luminosity $L_{\star}=2.2\times10^{10}\,\mathrm{L}_{\odot}$, agrees with typical halo mass to luminosity relations \citep[e.g.][]{vale04}. Finally, the median redshifts for the multiple image systems optimized in this model, i.e. all the new ones identified in UNCOVER imaging of the extended structures, are given in Tab.~\ref{tab:multiple_images1}. Note that in the case of multiple image systems that split into several sub-systems, we report the median redshift of the best reproduced sub-system.

We make our SL model publicly available on the UNCOVER website under: \url{https://jwst-uncover.github.io/DR1.html#LensingMaps}. The release comprises $7.6\arcmin\times7.6\arcmin$ deflection maps, $\alpha_x$ and $\alpha_y$, a convergence $\kappa$-map, and shear $\gamma_1$- and $\gamma_2$-maps, which cover the entire UNCOVER NIRCam footprint. We provide maps both in NIRCam UNCOVER pixel scale of $0.04\arcsec/\mathrm{pix}$, and in lower $0.1\arcsec/\mathrm{pix}$ resolution for the best-fit model and a range of low-resolution maps drawn from the final MCMC chain for error computation. The release also contains a set of magnification maps for various source redshifts.

The deep JWST imaging and SL mass model presented here for the first time fully reveal the SL power of the massive northern clumps ($\sim3\arcmin$ north and north-west of the main cluster core; see Fig.~\ref{fig:crit_curves}) in A2744. These were previously only constrained with WL measurements \citep{merten11,medezinski16,jauzac16b} which are much less sensitive to the small-scale mass distribution than the SL regime. While the WL analyses did indeed predict the three distinct northern sub-halos that we also model, we are now able to pinpoint their exact positions thanks to the more sensitive SL signal (see Tab.~\ref{tab:DM-halo_results}) and can confirm the halo positions to coincide with the three BCG-type galaxies identified in the north of A2744 (see section~\ref{sec:cluster_members}). Because of the different sensitivities of the SL and WL regimes, a direct quantitative comparison between the derived total masses is not possible. We nevertheless find the two northern sub-structures to be somewhat less massive than predicted by the WL, in particular by $\sim1$ order of magnitude compared to \citet{jauzac16b}. Note that all three WL studies predict an additional massive sub-halo west of the cluster core \citep{merten11,medezinski16,jauzac16b}, which lies just outside of the UNCOVER field-of-view. The extended HST data do not show a prominent cluster galaxy overdensity in that vicinity, even though there are two bright cluster galaxies at the position where \citet{merten11} predict the western sub-clump. We do not find any SL features in this region that could hint at an additional dense SL halo. That being said, the BUFFALO HST data in this region are shallower than the central HFF observations and possible faint SL features might therefore not be detected. Our model, however, does not require an external shear component (in early tries, introducing a freely optimized external shear free did not improve the model and the resulting external shear was very low), so if there is an additional sub-halo, it cannot be massive enough to significantly affect the SL signal of the multiple images identified in A2744 thus far. In addition, both \citet{medezinski16} and \citet{jauzac16b} predict another sub-structure in the north-east of the main cluster but disagree on its position. The UNCOVER NIRCam data reveal a small overdensity of apparent cluster galaxies a few arc-seconds from the \citet{jauzac16b} predicted centroid which seems to contain one prominent candidate galaxy-galaxy SL system. There is not sufficient HST coverage of this region however to robustly select cluster galaxies. This feature is therefore not included in the present analysis because we select the cluster galaxies based on their color around the 4000\,\AA-break in the HST data (see section~\ref{sec:cluster_members}). Future analyses should nevertheless investigate this system with spectroscopy to see if an additional sub-halo is needed in the north-east of the cluster. The most prominent sub-structure in A2744 by far is the north-western extension as can be seen in Figs.~\ref{fig:NW-extension}, \ref{fig:crit_curves} and~\ref{fig:magmap_z10}. We note that its position and morphology align well with the cosmic filament direction found by \citet{jauzac16b}. Future studies combining both WL and SL modeling will be able to further examine these conclusions and expand on them. Given the wide field-of-view and unparalleled spatial resolution of the UNCOVER imaging, it will be particularly interesting to see how an UNCOVER-based WL analysis would agree or disagree with our SL-based findings.

The main limitation of our model is that, unlike the secure multiple images around the main cluster core \citep{bergamini22}, the newly discovered images around the two northern sub-structures do not have spectroscopic redshifts, which makes the SL model somewhat less secure in that region. Indeed, low numbers of spectroscopic redshift constraints have been found to strongly affect e.g. magnification measurements \citep[e.g.][]{meneghetti17,acebron17,acebron18}. However, the multitude of systems together with the relatively tightly constrained photometric redshifts for some of them (see Tab.~\ref{tab:multiple_images1}) allow us to still obtain a relatively accurate solution as demonstrated by the low lens plane RMS achieved by our model. More robust modeling of these structures will however require spectroscopic redshift measurements of the new SL features detected in UNCOVER, either with VLT/MUSE for the brighter and bluer images or with JWST/NIRSpec for the fainter and redder ones such as our candidate $z\sim2.5$ overdensity (see section~\ref{sec:multiple_images}). The UNCOVER NIRSpec follow-up observations, scheduled for July~2023, for example will be of great use in this regard.

The newly established SL sub-structure in the north of A2744 also now allows us to more precisely estimate the magnification of background sources in and around the cluster field since these have been found to be significantly affected by massive structures in the outskirts of simulated lensing clusters \citep[e.g.][]{meneghetti17,acebron17}. Because of its unprecedented depth, large field-of-view and numerous ancillary data sets \citep[see][]{bezanson22}, the JWST UNCOVER field represents one of the prime fields for galaxy studies across the observable Universe. This is emphasized further by the relatively large high-magnification ($\mu\gtrsim4$) area spanned by the cluster sub-structures, in particular the main cluster and the north-western extension (see Fig.~\ref{fig:magmap_z10}), as discussed above. Studies of galaxies in this field will therefore require accurate gravitational magnification measurements such as provided by our SL model presented here and its subsequent versions.

\section{Conclusion} \label{sec:conculsion}
We present a new parametric SL model of the galaxy cluster A2744 based on hitherto unknown multiple image systems in the northern and north-western extensions of the cluster. Thanks to the ultra-deep and wide coverage of the JWST/NIRCam imaging taken for the UNCOVER program, we are able to identify numerous new multiple images and candidates in the two northern sub-structures of the cluster. Since no spectroscopic redshifts are available for these multiple image systems, we compile a relatively secure sub-set of SL constraints based on position, parity and photometric redshifts, as well as support by preliminary lens models. These new SL constraints are then combined with the \citet{bergamini22} spectroscopic sample and a now confirmed $z_{\mathrm{spec}}=9.76$ system \citep{roberts-borsani22} in the main cluster, forming a total sample of 135 multiple images used in our model. The SL model is constructed with an updated version of the parametric lensing mass modeling code by \citet{zitrin13a,zitrin13b,zitrin21} and comprises five large-scale cluster DM halos, five BCGs and a total of 421 cluster galaxies. Our final model reproduces the multiple images with a \emph{lens plane} RMS of 0.66\arcsec. 

Our main results are the following:

\begin{itemize}
    \item The UNCOVER imaging reveals two massive sub-structures ($\sim3\arcmin$) north and north-west of the central core of A2744. We uncover 75 new strongly lensed multiple images and multiple image candidates belonging to 17 sources around these structures, and establish them as prominent SL systems for the first time. These new multiple image systems include a potential overdensity of bright and red (i.e. HST-dark) sources at $z_{\mathrm{phot}}\sim2.5$ behind the north-western sub-structure. We use 43 of these new multiple images to constrain our SL model in these regions.
    \item Our model yields a total critical area of $A_{\mathrm{crit}}=0.67\,\mathrm{arcmin}^{2}$ for A2744 (assuming a source redshift $z_{\mathrm{s}}=2$) which implies an effective Einstein radius of $\theta_E=27.7\arcsec\pm2.8\arcsec$ enclosing a total projected mass of $M(<\theta_E)=(1.07\pm0.15)\times10^{14}\,\mathrm{M}_{\odot}$. This corresponds to a total cluster mass of $M(r<200\,\mathrm{kpc})=(1.60\pm0.22)\times10^{14}\,\mathrm{M}_{\odot}$ within 200\,kpc of BCG2.
    \item The main cluster core has an effective $z_{\mathrm{s}}=2$ Einstein radius $\theta_{E,\mathrm{main}}=23.2\arcsec\pm2.3\arcsec$ enclosing $M(<\theta_{E,\mathrm{main}})=(7.7\pm1.1)\times10^{13}\,\mathrm{M}_{\odot}$ and agrees well with previous SL models of A2744.
    \item The northern and north-western sub-clumps have effective $z_{\mathrm{s}}=2$ Einstein radii of $\theta_{E,\mathrm{N}}=7.4\arcsec\pm0.7\arcsec$ and $\theta_{E,\mathrm{NW}}=13.1\arcsec\pm1.3\arcsec$, enclosing total masses of $M(<\theta_{E,\mathrm{N}})=(0.8\pm0.1)\times10^{13}\,\mathrm{M}_{\odot}$ and $M(<\theta_{E,\mathrm{NW}})=(2.2\pm0.3)\times10^{13}\,\mathrm{M}_{\odot}$, respectively. These substructures are somewhat less massive than in previous SL studies in which these DM sub-halos were not directly constrained, however.
    \item Including the two northern sub-structures in the SL model effectively increased the critical area of the cluster by a factor $\sim1.5$. We in particular find an extended high-magnification (e.g. $\mu\gtrsim4$ for $z_{\mathrm{s}}=10$) area between the three cluster structures in the lens plane.
    \item Our SL-constrained sub-structures broadly agree with WL analysis results of A2744 but find much lower masses for the individual DM halos.
\end{itemize}

Our lens model is made available to the community, together with the public UNCOVER data release\footnote{\url{https://jwst-uncover.github.io/DR1.html}} \citep{bezanson22}. Future studies will need to measure spectroscopic redshifts of the newly discovered SL features in the outskirts of A2744 and include WL constraints to more robustly probe the complete DM sub-structure of the cluster. A part of that can be done with JWST UNCOVER data, including the upcoming NIRSpec spectroscopy. Our model will be updated in future versions with new constraints when they become available.

\section*{Acknowledgements}
We thank the anonymous referee for their useful comments and feedback that greatly helped to improve the completeness and clarity of this paper. LF and AZ would like to thank Ashish K. Meena for useful comments and discussions.

LF and AZ acknowledge support by Grant No. 2020750 from the United States-Israel Binational Science Foundation (BSF) and Grant No. 2109066 from the United States National Science Foundation (NSF). The BGU lensing group further acknowledges support by the Ministry of Science \& Technology, Israel. HA acknowledges support from CNES (Centre National d'Etudes Spatiales). RB acknowledges support from the Research Corporation for Scientific Advancement (RCSA) Cottrell Scholar Award ID No.: 27587. PD acknowledges support from the NWO grant 016.VIDI.189.162 (``ODIN'') and from the European Commission's and University of Groningen's CO-FUND Rosalind Franklin program. The Cosmic Dawn Center is funded by the Danish National Research Foundation (DNRF) under grant \#140. Support for the program JWST-GO-02561 was provided through a grant from the STScI under NASA contract NAS 5-03127.

This work is based on observations obtained with the NASA/ESA/CSA JWST and the NASA/ESA \textit{Hubble Space Telescope} (HST), retrieved from the \texttt{Barbara A. Mikulski Archive for Space Telescopes} (\texttt{MAST}) at the \textit{Space Telescope Science Institute} (STScI). STScI is operated by the Association of Universities for Research in Astronomy, Inc. under NASA contract NAS 5-26555. This work is also based on observations made with ESO Telescopes at the La Silla Paranal Observatory which are publicly available on the ESO Science Archive Facility. 

This research made use of \texttt{Astropy},\footnote{\url{http://www.astropy.org}} a community-developed core Python package for Astronomy \citep{astropy13,astropy18} as well as the packages \texttt{NumPy} \citep{vanderwalt11}, \texttt{SciPy} \citep{virtanen20}, \texttt{Matplotlib} \citep{hunter07} and the \texttt{MAAT} Astronomy and Astrophysics tools for \texttt{MATLAB} \citep[][]{maat14}.


\section*{Data Availability}
The raw JWST and HST data used in this work are publicly available on the \texttt{MAST} and the reduced UNCOVER JWST and HST mosaics and catalogs are part of the public data release by the UNCOVER team\footnote{\url{https://jwst-uncover.github.io}} \citep[][]{bezanson22,weaver23}. The spectroscopic redshifts were measured by \citet{mahler18}, \citet{bergamini22} and \citet{roberts-borsani22} based on data that are publicly available on the ESO Science Archive\footnote{\url{https://archive.eso.org/scienceportal/home}} and the GLASS website\footnote{\url{https://glass.astro.ucla.edu}}.


\bibliographystyle{mnras}
\bibliography{references} 

\begin{thebibliography}{}
\makeatletter
\relax
\def\mn@urlcharsother{\let\do\@makeother \do\$\do\&\do\#\do\^\do\_\do\%\do\~}
\def\mn@doi{\begingroup\mn@urlcharsother \@ifnextchar [ {\mn@doi@}
  {\mn@doi@[]}}
\def\mn@doi@[#1]#2{\def\@tempa{#1}\ifx\@tempa\@empty \href
  {http://dx.doi.org/#2} {doi:#2}\else \href {http://dx.doi.org/#2} {#1}\fi
  \endgroup}
\def\mn@eprint#1#2{\mn@eprint@#1:#2::\@nil}
\def\mn@eprint@arXiv#1{\href {http://arxiv.org/abs/#1} {{\tt arXiv:#1}}}
\def\mn@eprint@dblp#1{\href {http://dblp.uni-trier.de/rec/bibtex/#1.xml}
  {dblp:#1}}
\def\mn@eprint@#1:#2:#3:#4\@nil{\def\@tempa {#1}\def\@tempb {#2}\def\@tempc
  {#3}\ifx \@tempc \@empty \let \@tempc \@tempb \let \@tempb \@tempa \fi \ifx
  \@tempb \@empty \def\@tempb {arXiv}\fi \@ifundefined
  {mn@eprint@\@tempb}{\@tempb:\@tempc}{\expandafter \expandafter \csname
  mn@eprint@\@tempb\endcsname \expandafter{\@tempc}}}

\bibitem[\protect\citeauthoryear{{Abell}, {Corwin}  \& {Olowin}}{{Abell}
  et~al.}{1989}]{abell89}
{Abell} G.~O.,  {Corwin} Jr. H.~G.,   {Olowin} R.~P.,  1989, \mn@doi [\apjs]
  {10.1086/191333}, \href {http://cdsads.u-strasbg.fr/abs/1989ApJS...70....1A}
  {70, 1}

\bibitem[\protect\citeauthoryear{{Acebron}, {Jullo}, {Limousin}, {Tilquin},
  {Giocoli}, {Jauzac}, {Mahler}  \& {Richard}}{{Acebron}
  et~al.}{2017}]{acebron17}
{Acebron} A.,  {Jullo} E.,  {Limousin} M.,  {Tilquin} A.,  {Giocoli} C.,
  {Jauzac} M.,  {Mahler} G.,   {Richard} J.,  2017, \mn@doi [\mnras]
  {10.1093/mnras/stx1330}, \href
  {https://ui.adsabs.harvard.edu/abs/2017MNRAS.470.1809A} {470, 1809}

\bibitem[\protect\citeauthoryear{{Acebron} et~al.,}{{Acebron}
  et~al.}{2018}]{acebron18}
{Acebron} A.,  et~al., 2018, \mn@doi [\apj] {10.3847/1538-4357/aabe29}, \href
  {https://ui.adsabs.harvard.edu/abs/2018ApJ...858...42A} {858, 42}

\bibitem[\protect\citeauthoryear{{Allen}}{{Allen}}{1998}]{allen98}
{Allen} S.~W.,  1998, \mn@doi [\mnras] {10.1046/j.1365-8711.1998.01358.x},
  \href {https://ui.adsabs.harvard.edu/abs/1998MNRAS.296..392A} {296, 392}

\bibitem[\protect\citeauthoryear{{Amoura}, {Drakos}, {Berrouet}  \&
  {Taylor}}{{Amoura} et~al.}{2021}]{amoura21}
{Amoura} Y.,  {Drakos} N.~E.,  {Berrouet} A.,   {Taylor} J.~E.,  2021, \mn@doi
  [\mnras] {10.1093/mnras/stab2467}, \href
  {https://ui.adsabs.harvard.edu/abs/2021MNRAS.508..100A} {508, 100}

\bibitem[\protect\citeauthoryear{{Asencio}, {Banik}  \& {Kroupa}}{{Asencio}
  et~al.}{2021}]{asencio21}
{Asencio} E.,  {Banik} I.,   {Kroupa} P.,  2021, \mn@doi [\mnras]
  {10.1093/mnras/staa3441}, \href
  {https://ui.adsabs.harvard.edu/abs/2021MNRAS.500.5249A} {500, 5249}

\bibitem[\protect\citeauthoryear{{Astropy Collaboration} et~al.,}{{Astropy
  Collaboration} et~al.}{2013}]{astropy13}
{Astropy Collaboration} et~al., 2013, \mn@doi [\aap]
  {10.1051/0004-6361/201322068}, \href
  {http://adsabs.harvard.edu/abs/2013A%26A...558A..33A} {558, A33}

\bibitem[\protect\citeauthoryear{{Ata}, {Lee}, {Vecchia}, {Kitaura},
  {Cucciati}, {Lemaux}, {Kashino}  \& {M{\"u}ller}}{{Ata} et~al.}{2022}]{ata22}
{Ata} M.,  {Lee} K.-G.,  {Vecchia} C.~D.,  {Kitaura} F.-S.,  {Cucciati} O.,
  {Lemaux} B.~C.,  {Kashino} D.,   {M{\"u}ller} T.,  2022, \mn@doi [Nature
  Astronomy] {10.1038/s41550-022-01693-0}, \href
  {https://ui.adsabs.harvard.edu/abs/2022NatAs...6..857A} {6, 857}

\bibitem[\protect\citeauthoryear{{Atek} et~al.,}{{Atek} et~al.}{2014}]{atek14a}
{Atek} H.,  et~al., 2014, \mn@doi [\apj] {10.1088/0004-637X/786/1/60}, \href
  {http://adsabs.harvard.edu/abs/2014ApJ...786...60A} {786, 60}

\bibitem[\protect\citeauthoryear{{Atek} et~al.,}{{Atek} et~al.}{2015}]{atek15a}
{Atek} H.,  et~al., 2015, \mn@doi [\apj] {10.1088/0004-637X/800/1/18}, \href
  {http://cdsads.u-strasbg.fr/abs/2015ApJ...800...18A} {800, 18}

\bibitem[\protect\citeauthoryear{{Atek}, {Richard}, {Kneib}  \&
  {Schaerer}}{{Atek} et~al.}{2018}]{atek18}
{Atek} H.,  {Richard} J.,  {Kneib} J.-P.,   {Schaerer} D.,  2018, \mn@doi
  [\mnras] {10.1093/mnras/sty1820}, \href
  {https://ui.adsabs.harvard.edu/abs/2018MNRAS.479.5184A} {479, 5184}

\bibitem[\protect\citeauthoryear{{Bacon} et~al.,}{{Bacon}
  et~al.}{2010}]{bacon10}
{Bacon} R.,  et~al., 2010, in {McLean} I.~S.,  {Ramsay} S.~K.,   {Takami} H.,
  eds,  Society of Photo-Optical Instrumentation Engineers (SPIE) Conference
  Series Vol. 7735, Ground-based and Airborne Instrumentation for Astronomy
  III. p. 773508, \mn@doi{10.1117/12.856027}

\bibitem[\protect\citeauthoryear{Barbary}{Barbary}{2016}]{barbary16}
Barbary K.,  2016, \mn@doi [Journal of Open Source Software]
  {10.21105/joss.00058}, 1, 58

\bibitem[\protect\citeauthoryear{{Barbary}}{{Barbary}}{2018}]{barabry18}
{Barbary} K.,  2018, {SEP: Source Extraction and Photometry}, Astrophysics
  Source Code Library, record ascl:1811.004 (\mn@eprint {ascl} {1811.004})

\bibitem[\protect\citeauthoryear{{Barrufet} et~al.,}{{Barrufet}
  et~al.}{2023}]{barrufet22}
{Barrufet} L.,  et~al., 2023, \mn@doi [\mnras] {10.1093/mnras/stad947}, \href
  {https://ui.adsabs.harvard.edu/abs/2023MNRAS.522..449B} {522, 449}

\bibitem[\protect\citeauthoryear{{Bergamini} et~al.,}{{Bergamini}
  et~al.}{2019}]{bergamini19}
{Bergamini} P.,  et~al., 2019, \mn@doi [\aap] {10.1051/0004-6361/201935974},
  \href {https://ui.adsabs.harvard.edu/abs/2019A&A...631A.130B} {631, A130}

\bibitem[\protect\citeauthoryear{{Bergamini} et~al.,}{{Bergamini}
  et~al.}{2021}]{bergamini21}
{Bergamini} P.,  et~al., 2021, \mn@doi [\aap] {10.1051/0004-6361/202039564},
  \href {https://ui.adsabs.harvard.edu/abs/2021A&A...645A.140B} {645, A140}

\bibitem[\protect\citeauthoryear{{Bergamini} et~al.,}{{Bergamini}
  et~al.}{2022}]{bergamini22}
{Bergamini} P.,  et~al., 2022, arXiv e-prints, \href
  {https://ui.adsabs.harvard.edu/abs/2022arXiv220709416B} {p. arXiv:2207.09416}

\bibitem[\protect\citeauthoryear{Bertin \& Arnouts}{Bertin \&
  Arnouts}{1996}]{bertin96}
Bertin E.,  Arnouts S.,  1996, \mn@doi [Astronomy and Astrophysics Supplement
  Series] {10.1051/aas:1996164}, 117, 393

\bibitem[\protect\citeauthoryear{{Bezanson} et~al.,}{{Bezanson}
  et~al.}{2022}]{bezanson22}
{Bezanson} R.,  et~al., 2022, arXiv e-prints, \href
  {https://ui.adsabs.harvard.edu/abs/2022arXiv221204026B} {p. arXiv:2212.04026}

\bibitem[\protect\citeauthoryear{{Bhatawdekar}, {Conselice},
  {Margalef-Bentabol}  \& {Duncan}}{{Bhatawdekar} et~al.}{2019}]{bhatawdekar19}
{Bhatawdekar} R.,  {Conselice} C.~J.,  {Margalef-Bentabol} B.,   {Duncan} K.,
  2019, \mn@doi [\mnras] {10.1093/mnras/stz866}, \href
  {https://ui.adsabs.harvard.edu/abs/2019MNRAS.486.3805B} {486, 3805}

\bibitem[\protect\citeauthoryear{{B{\"o}ker} et~al.,}{{B{\"o}ker}
  et~al.}{2023}]{boeker23}
{B{\"o}ker} T.,  et~al., 2023, \mn@doi [\pasp] {10.1088/1538-3873/acb846},
  \href {https://ui.adsabs.harvard.edu/abs/2023PASP..135c8001B} {135, 038001}

\bibitem[\protect\citeauthoryear{{Borgani} \& {Guzzo}}{{Borgani} \&
  {Guzzo}}{2001}]{borgani01}
{Borgani} S.,  {Guzzo} L.,  2001, \mn@doi [\nat] {10.1038/409039A0}, \href
  {https://ui.adsabs.harvard.edu/abs/2001Natur.409...39B} {409, 39}

\bibitem[\protect\citeauthoryear{{Bouwens}, {Oesch}, {Illingworth}, {Ellis}  \&
  {Stefanon}}{{Bouwens} et~al.}{2017}]{bouwens17}
{Bouwens} R.~J.,  {Oesch} P.~A.,  {Illingworth} G.~D.,  {Ellis} R.~S.,
  {Stefanon} M.,  2017, \mn@doi [\apj] {10.3847/1538-4357/aa70a4}, \href
  {http://adsabs.harvard.edu/abs/2017ApJ...843..129B} {843, 129}

\bibitem[\protect\citeauthoryear{{Bouwens}, {Illingworth}, {Ellis}, {Oesch},
  {Paulino-Afonso}, {Ribeiro}  \& {Stefanon}}{{Bouwens}
  et~al.}{2022}]{bouwens22a}
{Bouwens} R.~J.,  {Illingworth} G.,  {Ellis} R.~S.,  {Oesch} P.,
  {Paulino-Afonso} A.,  {Ribeiro} B.,   {Stefanon} M.,  2022, \mn@doi [\apj]
  {10.3847/1538-4357/ac618c}, \href
  {https://ui.adsabs.harvard.edu/abs/2022ApJ...931...81B} {931, 81}

\bibitem[\protect\citeauthoryear{{Braglia}, {Pierini}, {Biviano}  \&
  {Boehringer}}{{Braglia} et~al.}{2009}]{braglia09}
{Braglia} F.~G.,  {Pierini} D.,  {Biviano} A.,   {Boehringer} H.,  2009, VizieR
  Online Data Catalog, \href
  {https://ui.adsabs.harvard.edu/abs/2009yCat..35000947B} {pp J/A+A/500/947}

\bibitem[\protect\citeauthoryear{{Brammer}, {van Dokkum}  \& {Coppi}}{{Brammer}
  et~al.}{2008}]{brammer08}
{Brammer} G.~B.,  {van Dokkum} P.~G.,   {Coppi} P.,  2008, \mn@doi [\apj]
  {10.1086/591786}, \href
  {https://ui.adsabs.harvard.edu/abs/2008ApJ...686.1503B} {686, 1503}

\bibitem[\protect\citeauthoryear{{Broadhurst} \& {Barkana}}{{Broadhurst} \&
  {Barkana}}{2008}]{broadhurst08b}
{Broadhurst} T.~J.,  {Barkana} R.,  2008, \mn@doi [\mnras]
  {10.1111/j.1365-2966.2008.13852.x}, \href
  {https://ui.adsabs.harvard.edu/abs/2008MNRAS.390.1647B} {390, 1647}

\bibitem[\protect\citeauthoryear{{Broadhurst}, {Umetsu}, {Medezinski}, {Oguri}
  \& {Rephaeli}}{{Broadhurst} et~al.}{2008}]{broadhurst08a}
{Broadhurst} T.,  {Umetsu} K.,  {Medezinski} E.,  {Oguri} M.,   {Rephaeli} Y.,
  2008, \mn@doi [\apjl] {10.1086/592400}, \href
  {https://ui.adsabs.harvard.edu/abs/2008ApJ...685L...9B} {685, L9}

\bibitem[\protect\citeauthoryear{{Caminha}, {Suyu}, {Mercurio}, {Brammer},
  {Bergamini}, {Acebron}  \& {Vanzella}}{{Caminha} et~al.}{2022}]{caminha22}
{Caminha} G.~B.,  {Suyu} S.~H.,  {Mercurio} A.,  {Brammer} G.,  {Bergamini} P.,
   {Acebron} A.,   {Vanzella} E.,  2022, \mn@doi [\aap]
  {10.1051/0004-6361/202244517}, \href
  {https://ui.adsabs.harvard.edu/abs/2022A&A...666L...9C} {666, L9}

\bibitem[\protect\citeauthoryear{{Coe}, {Bradley}  \& {Zitrin}}{{Coe}
  et~al.}{2015}]{coe15}
{Coe} D.,  {Bradley} L.,   {Zitrin} A.,  2015, \mn@doi [\apj]
  {10.1088/0004-637X/800/2/84}, \href
  {https://ui.adsabs.harvard.edu/abs/2015ApJ...800...84C} {800, 84}

\bibitem[\protect\citeauthoryear{{Coe} et~al.,}{{Coe} et~al.}{2019}]{coe19}
{Coe} D.,  et~al., 2019, \mn@doi [\apj] {10.3847/1538-4357/ab412b}, \href
  {https://ui.adsabs.harvard.edu/abs/2019ApJ...884...85C} {884, 85}

\bibitem[\protect\citeauthoryear{{Diego} et~al.,}{{Diego}
  et~al.}{2023}]{diego22}
{Diego} J.~M.,  et~al., 2023, \mn@doi [\aap] {10.1051/0004-6361/202245238},
  \href {https://ui.adsabs.harvard.edu/abs/2023A&A...672A...3D} {672, A3}

\bibitem[\protect\citeauthoryear{{Doyon} et~al.,}{{Doyon}
  et~al.}{2012}]{doyon12}
{Doyon} R.,  et~al., 2012, in {Clampin} M.~C.,  {Fazio} G.~G.,  {MacEwen}
  H.~A.,   {Oschmann} Jacobus~M. J.,  eds,  Society of Photo-Optical
  Instrumentation Engineers (SPIE) Conference Series Vol. 8442, Space
  Telescopes and Instrumentation 2012: Optical, Infrared, and Millimeter Wave.
  p. 84422R, \mn@doi{10.1117/12.926578}

\bibitem[\protect\citeauthoryear{{Dubois} et~al.,}{{Dubois}
  et~al.}{2014}]{dubois14}
{Dubois} Y.,  et~al., 2014, \mn@doi [\mnras] {10.1093/mnras/stu1227}, \href
  {https://ui.adsabs.harvard.edu/abs/2014MNRAS.444.1453D} {444, 1453}

\bibitem[\protect\citeauthoryear{{Durret} et~al.,}{{Durret}
  et~al.}{2016}]{durret16}
{Durret} F.,  et~al., 2016, \mn@doi [\aap] {10.1051/0004-6361/201527655}, \href
  {https://ui.adsabs.harvard.edu/abs/2016A&A...588A..69D} {588, A69}

\bibitem[\protect\citeauthoryear{{Ebeling}, {Barrett}  \& {Donovan}}{{Ebeling}
  et~al.}{2004}]{ebeling04}
{Ebeling} H.,  {Barrett} E.,   {Donovan} D.,  2004, \mn@doi [\apjl]
  {10.1086/422750}, \href
  {https://ui.adsabs.harvard.edu/abs/2004ApJ...609L..49E} {609, L49}

\bibitem[\protect\citeauthoryear{{Ebeling}, {Barrett}, {Donovan}, {Ma}, {Edge}
  \& {van Speybroeck}}{{Ebeling} et~al.}{2007}]{ebeling07}
{Ebeling} H.,  {Barrett} E.,  {Donovan} D.,  {Ma} C.-J.,  {Edge} A.~C.,   {van
  Speybroeck} L.,  2007, \mn@doi [\apjl] {10.1086/518603}, \href
  {http://adsabs.harvard.edu/abs/2007ApJ...661L..33E} {661, L33}

\bibitem[\protect\citeauthoryear{{Ebeling}, {Edge}, {Mantz}, {Barrett},
  {Henry}, {Ma}  \& {van Speybroeck}}{{Ebeling} et~al.}{2010}]{ebeling10}
{Ebeling} H.,  {Edge} A.~C.,  {Mantz} A.,  {Barrett} E.,  {Henry} J.~P.,  {Ma}
  C.~J.,   {van Speybroeck} L.,  2010, \mn@doi [\mnras]
  {10.1111/j.1365-2966.2010.16920.x}, \href
  {http://adsabs.harvard.edu/abs/2010MNRAS.407...83E} {407, 83}

\bibitem[\protect\citeauthoryear{{El{\'{\i}}asd{\'o}ttir}
  et~al.,}{{El{\'{\i}}asd{\'o}ttir} et~al.}{2007}]{eliasdottir07}
{El{\'{\i}}asd{\'o}ttir} {\'A}.,  et~al., 2007, preprint, \href
  {http://cdsads.u-strasbg.fr/abs/2007arXiv0710.5636E} {} (\mn@eprint {arXiv}
  {0710.5636})

\bibitem[\protect\citeauthoryear{{Ellien}, {Durret}, {Adami}, {Martinet},
  {Lobo}  \& {Jauzac}}{{Ellien} et~al.}{2019}]{ellien19}
{Ellien} A.,  {Durret} F.,  {Adami} C.,  {Martinet} N.,  {Lobo} C.,   {Jauzac}
  M.,  2019, \mn@doi [\aap] {10.1051/0004-6361/201935673}, \href
  {https://ui.adsabs.harvard.edu/abs/2019A&A...628A..34E} {628, A34}

\bibitem[\protect\citeauthoryear{{Faber} \& {Jackson}}{{Faber} \&
  {Jackson}}{1976}]{faber-jackson76}
{Faber} S.~M.,  {Jackson} R.~E.,  1976, \mn@doi [\apj] {10.1086/154215}, \href
  {https://ui.adsabs.harvard.edu/abs/1976ApJ...204..668F} {204, 668}

\bibitem[\protect\citeauthoryear{{Ferruit} et~al.,}{{Ferruit}
  et~al.}{2022}]{ferruit22}
{Ferruit} P.,  et~al., 2022, \mn@doi [\aap] {10.1051/0004-6361/202142673},
  \href {https://ui.adsabs.harvard.edu/abs/2022A&A...661A..81F} {661, A81}

\bibitem[\protect\citeauthoryear{{Foley} et~al.,}{{Foley}
  et~al.}{2011}]{foley11}
{Foley} R.~J.,  et~al., 2011, \mn@doi [\apj] {10.1088/0004-637X/731/2/86},
  \href {https://ui.adsabs.harvard.edu/abs/2011ApJ...731...86F} {731, 86}

\bibitem[\protect\citeauthoryear{{Fudamoto}, {Inoue}  \& {Sugahara}}{{Fudamoto}
  et~al.}{2022}]{fudamoto22}
{Fudamoto} Y.,  {Inoue} A.~K.,   {Sugahara} Y.,  2022, arXiv e-prints, \href
  {https://ui.adsabs.harvard.edu/abs/2022arXiv220800132F} {p. arXiv:2208.00132}

\bibitem[\protect\citeauthoryear{{Furtak}, {Atek}, {Lehnert}, {Chevallard}  \&
  {Charlot}}{{Furtak} et~al.}{2021}]{furtak21}
{Furtak} L.~J.,  {Atek} H.,  {Lehnert} M.~D.,  {Chevallard} J.,   {Charlot} S.,
   2021, \mn@doi [\mnras] {10.1093/mnras/staa3760}, \href
  {https://ui.adsabs.harvard.edu/abs/2021MNRAS.501.1568F} {501, 1568}

\bibitem[\protect\citeauthoryear{{Furtak} et~al.,}{{Furtak}
  et~al.}{2022}]{furtak22b}
{Furtak} L.~J.,  et~al., 2022, \mn@doi [arXiv e-prints]
  {10.48550/arXiv.2212.10531}, \href
  {https://ui.adsabs.harvard.edu/abs/2022arXiv221210531F} {p. arXiv:2212.10531}

\bibitem[\protect\citeauthoryear{{Gardner} et~al.,}{{Gardner}
  et~al.}{2006}]{gardner06}
{Gardner} J.~P.,  et~al., 2006, {Science with the James Webb space telescope}.
SPIE, p. 62650N, \mn@doi{10.1117/12.670492}

\bibitem[\protect\citeauthoryear{{Glazebrook} et~al.,}{{Glazebrook}
  et~al.}{2023}]{glazebrook22}
{Glazebrook} K.,  et~al., 2023, \mn@doi [\apjl] {10.3847/2041-8213/acba8b},
  \href {https://ui.adsabs.harvard.edu/abs/2023ApJ...947L..25G} {947, L25}

\bibitem[\protect\citeauthoryear{{Gonz{\'a}lez-L{\'o}pez}
  et~al.,}{{Gonz{\'a}lez-L{\'o}pez} et~al.}{2017}]{gonzales-lopez17}
{Gonz{\'a}lez-L{\'o}pez} J.,  et~al., 2017, \mn@doi [\aap]
  {10.1051/0004-6361/201628806}, \href
  {https://ui.adsabs.harvard.edu/abs/2017A&A...597A..41G} {597, A41}

\bibitem[\protect\citeauthoryear{{Grillo} et~al.,}{{Grillo}
  et~al.}{2015}]{grillo15}
{Grillo} C.,  et~al., 2015, \mn@doi [\apj] {10.1088/0004-637X/800/1/38}, \href
  {http://adsabs.harvard.edu/abs/2015ApJ...800...38G} {800, 38}

\bibitem[\protect\citeauthoryear{{Halkola}, {Seitz}  \& {Pannella}}{{Halkola}
  et~al.}{2006}]{halkola06}
{Halkola} A.,  {Seitz} S.,   {Pannella} M.,  2006, \mn@doi [\mnras]
  {10.1111/j.1365-2966.2006.10948.x}, \href
  {http://adsabs.harvard.edu/abs/2006MNRAS.372.1425H} {372, 1425}

\bibitem[\protect\citeauthoryear{{Halkola}, {Seitz}  \& {Pannella}}{{Halkola}
  et~al.}{2007}]{halkola07}
{Halkola} A.,  {Seitz} S.,   {Pannella} M.,  2007, \mn@doi [\apj]
  {10.1086/510555}, \href
  {https://ui.adsabs.harvard.edu/abs/2007ApJ...656..739H} {656, 739}

\bibitem[\protect\citeauthoryear{{Hatfield}, {Laigle}, {Jarvis}, {Devriendt},
  {Davidzon}, {Ilbert}, {Pichon}  \& {Dubois}}{{Hatfield}
  et~al.}{2019}]{hatfield19}
{Hatfield} P.~W.,  {Laigle} C.,  {Jarvis} M.~J.,  {Devriendt} J.,  {Davidzon}
  I.,  {Ilbert} O.,  {Pichon} C.,   {Dubois} Y.,  2019, \mn@doi [\mnras]
  {10.1093/mnras/stz2946}, \href
  {https://ui.adsabs.harvard.edu/abs/2019MNRAS.490.5043H} {490, 5043}

\bibitem[\protect\citeauthoryear{{Hennawi}, {Dalal}, {Bode}  \&
  {Ostriker}}{{Hennawi} et~al.}{2007}]{hennawi07}
{Hennawi} J.~F.,  {Dalal} N.,  {Bode} P.,   {Ostriker} J.~P.,  2007, \mn@doi
  [\apj] {10.1086/497362}, \href
  {https://ui.adsabs.harvard.edu/abs/2007ApJ...654..714H} {654, 714}

\bibitem[\protect\citeauthoryear{{Hsiao} et~al.,}{{Hsiao}
  et~al.}{2022}]{hsiao22}
{Hsiao} T. Y.-Y.,  et~al., 2022, arXiv e-prints, \href
  {https://ui.adsabs.harvard.edu/abs/2022arXiv221014123H} {p. arXiv:2210.14123}

\bibitem[\protect\citeauthoryear{Hunter}{Hunter}{2007}]{hunter07}
Hunter J.~D.,  2007, \mn@doi [Computing in Science \& Engineering]
  {10.1109/MCSE.2007.55}, 9, 90

\bibitem[\protect\citeauthoryear{{Ishigaki}, {Kawamata}, {Ouchi}, {Oguri},
  {Shimasaku}  \& {Ono}}{{Ishigaki} et~al.}{2015}]{ishigaki15a}
{Ishigaki} M.,  {Kawamata} R.,  {Ouchi} M.,  {Oguri} M.,  {Shimasaku} K.,
  {Ono} Y.,  2015, \mn@doi [\apj] {10.1088/0004-637X/799/1/12}, \href
  {http://cdsads.u-strasbg.fr/abs/2015ApJ...799...12I} {799, 12}

\bibitem[\protect\citeauthoryear{{Ishigaki}, {Kawamata}, {Ouchi}, {Oguri},
  {Shimasaku}  \& {Ono}}{{Ishigaki} et~al.}{2018}]{ishigaki18}
{Ishigaki} M.,  {Kawamata} R.,  {Ouchi} M.,  {Oguri} M.,  {Shimasaku} K.,
  {Ono} Y.,  2018, \mn@doi [\apj] {10.3847/1538-4357/aaa544}, \href
  {http://adsabs.harvard.edu/abs/2018ApJ...854...73I} {854, 73}

\bibitem[\protect\citeauthoryear{{Jaffe}}{{Jaffe}}{1983}]{jaffe83}
{Jaffe} W.,  1983, \mn@doi [\mnras] {10.1093/mnras/202.4.995}, \href
  {https://ui.adsabs.harvard.edu/abs/1983MNRAS.202..995J} {202, 995}

\bibitem[\protect\citeauthoryear{{Jakobsen} et~al.,}{{Jakobsen}
  et~al.}{2022}]{jakobsen22}
{Jakobsen} P.,  et~al., 2022, \mn@doi [\aap] {10.1051/0004-6361/202142663},
  \href {https://ui.adsabs.harvard.edu/abs/2022A&A...661A..80J} {661, A80}

\bibitem[\protect\citeauthoryear{{Jauzac} et~al.,}{{Jauzac}
  et~al.}{2012}]{jauzac12}
{Jauzac} M.,  et~al., 2012, \mn@doi [\mnras]
  {10.1111/j.1365-2966.2012.21966.x}, \href
  {https://ui.adsabs.harvard.edu/abs/2012MNRAS.426.3369J} {426, 3369}

\bibitem[\protect\citeauthoryear{{Jauzac} et~al.,}{{Jauzac}
  et~al.}{2015}]{jauzac15}
{Jauzac} M.,  et~al., 2015, \mn@doi [\mnras] {10.1093/mnras/stv1402}, \href
  {http://cdsads.u-strasbg.fr/abs/2015MNRAS.452.1437J} {452, 1437}

\bibitem[\protect\citeauthoryear{{Jauzac} et~al.,}{{Jauzac}
  et~al.}{2016}]{jauzac16b}
{Jauzac} M.,  et~al., 2016, \mn@doi [\mnras] {10.1093/mnras/stw2251}, \href
  {https://ui.adsabs.harvard.edu/abs/2016MNRAS.463.3876J} {463, 3876}

\bibitem[\protect\citeauthoryear{{Jauzac} et~al.,}{{Jauzac}
  et~al.}{2018}]{jauzac18}
{Jauzac} M.,  et~al., 2018, \mn@doi [\mnras] {10.1093/mnras/sty2366}, \href
  {https://ui.adsabs.harvard.edu/abs/2018MNRAS.481.2901J} {481, 2901}

\bibitem[\protect\citeauthoryear{{Johnson} \& {Sharon}}{{Johnson} \&
  {Sharon}}{2016}]{johnson16}
{Johnson} T.~L.,  {Sharon} K.,  2016, \mn@doi [\apj]
  {10.3847/0004-637X/832/1/82}, \href
  {https://ui.adsabs.harvard.edu/abs/2016ApJ...832...82J} {832, 82}

\bibitem[\protect\citeauthoryear{{Johnson}, {Sharon}, {Bayliss}, {Gladders},
  {Coe}  \& {Ebeling}}{{Johnson} et~al.}{2014}]{johnson14}
{Johnson} T.~L.,  {Sharon} K.,  {Bayliss} M.~B.,  {Gladders} M.~D.,  {Coe} D.,
   {Ebeling} H.,  2014, \mn@doi [\apj] {10.1088/0004-637X/797/1/48}, \href
  {http://adsabs.harvard.edu/abs/2014ApJ...797...48J} {797, 48}

\bibitem[\protect\citeauthoryear{{Jullo} \& {Kneib}}{{Jullo} \&
  {Kneib}}{2009}]{jullo09}
{Jullo} E.,  {Kneib} J.-P.,  2009, \mn@doi [\mnras]
  {10.1111/j.1365-2966.2009.14654.x}, \href
  {http://cdsads.u-strasbg.fr/abs/2009MNRAS.395.1319J} {395, 1319}

\bibitem[\protect\citeauthoryear{{Jullo}, {Kneib}, {Limousin},
  {El{\'{\i}}asd{\'o}ttir}, {Marshall}  \& {Verdugo}}{{Jullo}
  et~al.}{2007}]{jullo07}
{Jullo} E.,  {Kneib} J.-P.,  {Limousin} M.,  {El{\'{\i}}asd{\'o}ttir} {\'A}.,
  {Marshall} P.~J.,   {Verdugo} T.,  2007, \mn@doi [New Journal of Physics]
  {10.1088/1367-2630/9/12/447}, \href
  {http://adsabs.harvard.edu/abs/2007NJPh....9..447J} {9, 447}

\bibitem[\protect\citeauthoryear{{Kartaltepe}, {Ebeling}, {Ma}  \&
  {Donovan}}{{Kartaltepe} et~al.}{2008}]{kartaltepe08}
{Kartaltepe} J.~S.,  {Ebeling} H.,  {Ma} C.~J.,   {Donovan} D.,  2008, \mn@doi
  [\mnras] {10.1111/j.1365-2966.2008.13620.x}, \href
  {https://ui.adsabs.harvard.edu/abs/2008MNRAS.389.1240K} {389, 1240}

\bibitem[\protect\citeauthoryear{{Kawamata}, {Oguri}, {Ishigaki}, {Shimasaku}
  \& {Ouchi}}{{Kawamata} et~al.}{2016}]{kawamata16}
{Kawamata} R.,  {Oguri} M.,  {Ishigaki} M.,  {Shimasaku} K.,   {Ouchi} M.,
  2016, \mn@doi [\apj] {10.3847/0004-637X/819/2/114}, \href
  {https://ui.adsabs.harvard.edu/abs/2016ApJ...819..114K} {819, 114}

\bibitem[\protect\citeauthoryear{{Kawamata}, {Ishigaki}, {Shimasaku}, {Oguri},
  {Ouchi}  \& {Tanigawa}}{{Kawamata} et~al.}{2018}]{kawamata18}
{Kawamata} R.,  {Ishigaki} M.,  {Shimasaku} K.,  {Oguri} M.,  {Ouchi} M.,
  {Tanigawa} S.,  2018, \mn@doi [\apj] {10.3847/1538-4357/aaa6cf}, \href
  {https://ui.adsabs.harvard.edu/abs/2018ApJ...855....4K} {855, 4}

\bibitem[\protect\citeauthoryear{{Keeton}}{{Keeton}}{2001}]{keeton01a}
{Keeton} C.~R.,  2001, \mn@doi [arXiv e-prints]
  {10.48550/arXiv.astro-ph/0102341}, \href
  {https://ui.adsabs.harvard.edu/abs/2001astro.ph..2341K} {pp
  astro--ph/0102341}

\bibitem[\protect\citeauthoryear{{Keeton}}{{Keeton}}{2010}]{keeton10}
{Keeton} C.~R.,  2010, \mn@doi [General Relativity and Gravitation]
  {10.1007/s10714-010-1041-1}, \href
  {https://ui.adsabs.harvard.edu/abs/2010GReGr..42.2151K} {42, 2151}

\bibitem[\protect\citeauthoryear{{Kelly} et~al.,}{{Kelly}
  et~al.}{2015}]{kelly15}
{Kelly} P.~L.,  et~al., 2015, \mn@doi [Science] {10.1126/science.aaa3350},
  \href {https://ui.adsabs.harvard.edu/abs/2015Sci...347.1123K} {347, 1123}

\bibitem[\protect\citeauthoryear{{Kelly} et~al.,}{{Kelly}
  et~al.}{2016a}]{kelly16a}
{Kelly} P.~L.,  et~al., 2016a, \mn@doi [\apjl] {10.3847/2041-8205/819/1/L8},
  \href {https://ui.adsabs.harvard.edu/abs/2016ApJ...819L...8K} {819, L8}

\bibitem[\protect\citeauthoryear{{Kelly} et~al.,}{{Kelly}
  et~al.}{2016b}]{kelly16b}
{Kelly} P.~L.,  et~al., 2016b, \mn@doi [\apj] {10.3847/0004-637X/831/2/205},
  \href {https://ui.adsabs.harvard.edu/abs/2016ApJ...831..205K} {831, 205}

\bibitem[\protect\citeauthoryear{{Kelly} et~al.,}{{Kelly}
  et~al.}{2018}]{kelly18}
{Kelly} P.~L.,  et~al., 2018, \mn@doi [Nature Astronomy]
  {10.1038/s41550-018-0430-3}, \href
  {https://ui.adsabs.harvard.edu/abs/2018NatAs...2..334K} {2, 334}

\bibitem[\protect\citeauthoryear{{Kikuchihara} et~al.,}{{Kikuchihara}
  et~al.}{2020}]{kikuchihara20}
{Kikuchihara} S.,  et~al., 2020, \mn@doi [\apj] {10.3847/1538-4357/ab7dbe},
  \href {https://ui.adsabs.harvard.edu/abs/2020ApJ...893...60K} {893, 60}

\bibitem[\protect\citeauthoryear{{Kim}, {Jee}, {Perlmutter}, {Hayden}, {Rubin},
  {Huang}, {Aldering}  \& {Ko}}{{Kim} et~al.}{2019}]{kim19}
{Kim} J.,  {Jee} M.~J.,  {Perlmutter} S.,  {Hayden} B.,  {Rubin} D.,  {Huang}
  X.,  {Aldering} G.,   {Ko} J.,  2019, \mn@doi [\apj]
  {10.3847/1538-4357/ab521e}, \href
  {https://ui.adsabs.harvard.edu/abs/2019ApJ...887...76K} {887, 76}

\bibitem[\protect\citeauthoryear{{Klypin} \& {Shandarin}}{{Klypin} \&
  {Shandarin}}{1983}]{klypin83}
{Klypin} A.~A.,  {Shandarin} S.~F.,  1983, \mn@doi [\mnras]
  {10.1093/mnras/204.3.891}, \href
  {https://ui.adsabs.harvard.edu/abs/1983MNRAS.204..891K} {204, 891}

\bibitem[\protect\citeauthoryear{{Kneib} \& {Natarajan}}{{Kneib} \&
  {Natarajan}}{2011}]{kneib11}
{Kneib} J.-P.,  {Natarajan} P.,  2011, \mn@doi [\aapr]
  {10.1007/s00159-011-0047-3}, \href
  {https://ui.adsabs.harvard.edu/abs/2011A&ARv..19...47K} {19, 47}

\bibitem[\protect\citeauthoryear{{Kneib}, {Mellier}, {Fort}  \&
  {Mathez}}{{Kneib} et~al.}{1993}]{Kneib1993}
{Kneib} J.~P.,  {Mellier} Y.,  {Fort} B.,   {Mathez} G.,  1993, \aap, \href
  {https://ui.adsabs.harvard.edu/abs/1993A&A...273..367K} {273, 367}

\bibitem[\protect\citeauthoryear{{Kneib}, {Ellis}, {Smail}, {Couch}  \&
  {Sharples}}{{Kneib} et~al.}{1996}]{kneib96}
{Kneib} J.-P.,  {Ellis} R.~S.,  {Smail} I.,  {Couch} W.~J.,   {Sharples} R.~M.,
   1996, \mn@doi [\apj] {10.1086/177995}, \href
  {https://ui.adsabs.harvard.edu/abs/1996ApJ...471..643K} {471, 643}

\bibitem[\protect\citeauthoryear{{Livermore}, {Finkelstein}  \&
  {Lotz}}{{Livermore} et~al.}{2017}]{livermore17}
{Livermore} R.~C.,  {Finkelstein} S.~L.,   {Lotz} J.~M.,  2017, \mn@doi [\apj]
  {10.3847/1538-4357/835/2/113}, \href
  {http://adsabs.harvard.edu/abs/2017ApJ...835..113L} {835, 113}

\bibitem[\protect\citeauthoryear{{Lotz} et~al.,}{{Lotz} et~al.}{2017}]{lotz17}
{Lotz} J.~M.,  et~al., 2017, \mn@doi [\apj] {10.3847/1538-4357/837/1/97}, \href
  {http://adsabs.harvard.edu/abs/2017ApJ...837...97L} {837, 97}

\bibitem[\protect\citeauthoryear{{Mahler} et~al.,}{{Mahler}
  et~al.}{2018}]{mahler18}
{Mahler} G.,  et~al., 2018, \mn@doi [\mnras] {10.1093/mnras/stx1971}, \href
  {http://adsabs.harvard.edu/abs/2018MNRAS.473..663M} {473, 663}

\bibitem[\protect\citeauthoryear{{Mahler} et~al.,}{{Mahler}
  et~al.}{2023}]{mahler22}
{Mahler} G.,  et~al., 2023, \mn@doi [\apj] {10.3847/1538-4357/acaea9}, \href
  {https://ui.adsabs.harvard.edu/abs/2023ApJ...945...49M} {945, 49}

\bibitem[\protect\citeauthoryear{{Maizy}, {Richard}, {de Leo}, {Pell{\'o}}  \&
  {Kneib}}{{Maizy} et~al.}{2010}]{maizy10}
{Maizy} A.,  {Richard} J.,  {de Leo} M.~A.,  {Pell{\'o}} R.,   {Kneib} J.~P.,
  2010, \mn@doi [\aap] {10.1051/0004-6361/200911829}, \href
  {https://ui.adsabs.harvard.edu/abs/2010A&A...509A.105M} {509, A105}

\bibitem[\protect\citeauthoryear{{McElwain} et~al.,}{{McElwain}
  et~al.}{2023}]{mcelwain23}
{McElwain} M.~W.,  et~al., 2023, \mn@doi [arXiv e-prints]
  {10.48550/arXiv.2301.01779}, \href
  {https://ui.adsabs.harvard.edu/abs/2023arXiv230101779M} {p. arXiv:2301.01779}

\bibitem[\protect\citeauthoryear{{Medezinski}, {Umetsu}, {Okabe}, {Nonino},
  {Molnar}, {Massey}, {Dupke}  \& {Merten}}{{Medezinski}
  et~al.}{2016}]{medezinski16}
{Medezinski} E.,  {Umetsu} K.,  {Okabe} N.,  {Nonino} M.,  {Molnar} S.,
  {Massey} R.,  {Dupke} R.,   {Merten} J.,  2016, \mn@doi [\apj]
  {10.3847/0004-637X/817/1/24}, \href
  {https://ui.adsabs.harvard.edu/abs/2016ApJ...817...24M} {817, 24}

\bibitem[\protect\citeauthoryear{{Meena} et~al.,}{{Meena}
  et~al.}{2023a}]{meena22a}
{Meena} A.~K.,  et~al., 2023a, \mn@doi [\mnras] {10.1093/mnras/stad869}, \href
  {https://ui.adsabs.harvard.edu/abs/2023MNRAS.521.5224M} {521, 5224}

\bibitem[\protect\citeauthoryear{{Meena} et~al.,}{{Meena}
  et~al.}{2023b}]{meena22b}
{Meena} A.~K.,  et~al., 2023b, \mn@doi [\apjl] {10.3847/2041-8213/acb645},
  \href {https://ui.adsabs.harvard.edu/abs/2023ApJ...944L...6M} {944, L6}

\bibitem[\protect\citeauthoryear{{Meneghetti}, {Bartelmann}  \&
  {Moscardini}}{{Meneghetti} et~al.}{2003}]{meneghetti03}
{Meneghetti} M.,  {Bartelmann} M.,   {Moscardini} L.,  2003, \mn@doi [\mnras]
  {10.1046/j.1365-8711.2003.06276.x}, \href
  {https://ui.adsabs.harvard.edu/abs/2003MNRAS.340..105M} {340, 105}

\bibitem[\protect\citeauthoryear{{Meneghetti}, {Bartelmann}, {Jenkins}  \&
  {Frenk}}{{Meneghetti} et~al.}{2007a}]{meneghetti07a}
{Meneghetti} M.,  {Bartelmann} M.,  {Jenkins} A.,   {Frenk} C.,  2007a, \mn@doi
  [\mnras] {10.1111/j.1365-2966.2007.12225.x}, \href
  {https://ui.adsabs.harvard.edu/abs/2007MNRAS.381..171M} {381, 171}

\bibitem[\protect\citeauthoryear{{Meneghetti}, {Argazzi}, {Pace}, {Moscardini},
  {Dolag}, {Bartelmann}, {Li}  \& {Oguri}}{{Meneghetti}
  et~al.}{2007b}]{meneghetti07b}
{Meneghetti} M.,  {Argazzi} R.,  {Pace} F.,  {Moscardini} L.,  {Dolag} K.,
  {Bartelmann} M.,  {Li} G.,   {Oguri} M.,  2007b, \mn@doi [\aap]
  {10.1051/0004-6361:20065722}, \href
  {https://ui.adsabs.harvard.edu/abs/2007A&A...461...25M} {461, 25}

\bibitem[\protect\citeauthoryear{{Meneghetti} et~al.,}{{Meneghetti}
  et~al.}{2017}]{meneghetti17}
{Meneghetti} M.,  et~al., 2017, \mn@doi [\mnras] {10.1093/mnras/stx2064}, \href
  {https://ui.adsabs.harvard.edu/abs/2017MNRAS.472.3177M} {472, 3177}

\bibitem[\protect\citeauthoryear{{Merten} et~al.,}{{Merten}
  et~al.}{2011}]{merten11}
{Merten} J.,  et~al., 2011, \mn@doi [\mnras]
  {10.1111/j.1365-2966.2011.19266.x}, \href
  {http://adsabs.harvard.edu/abs/2011MNRAS.417..333M} {417, 333}

\bibitem[\protect\citeauthoryear{{Miller} et~al.,}{{Miller}
  et~al.}{2018}]{miller18}
{Miller} T.~B.,  et~al., 2018, \mn@doi [\nat] {10.1038/s41586-018-0025-2},
  \href {https://ui.adsabs.harvard.edu/abs/2018Natur.556..469M} {556, 469}

\bibitem[\protect\citeauthoryear{{Monna} et~al.,}{{Monna}
  et~al.}{2014}]{monna14}
{Monna} A.,  et~al., 2014, \mn@doi [\mnras] {10.1093/mnras/stt2284}, \href
  {https://ui.adsabs.harvard.edu/abs/2014MNRAS.438.1417M} {438, 1417}

\bibitem[\protect\citeauthoryear{{Monna} et~al.,}{{Monna}
  et~al.}{2015}]{monna15}
{Monna} A.,  et~al., 2015, \mn@doi [\mnras] {10.1093/mnras/stu2534}, \href
  {https://ui.adsabs.harvard.edu/abs/2015MNRAS.447.1224M} {447, 1224}

\bibitem[\protect\citeauthoryear{{Mullis}, {Rosati}, {Lamer}, {B{\"o}hringer},
  {Schwope}, {Schuecker}  \& {Fassbender}}{{Mullis} et~al.}{2005}]{mullis05}
{Mullis} C.~R.,  {Rosati} P.,  {Lamer} G.,  {B{\"o}hringer} H.,  {Schwope} A.,
  {Schuecker} P.,   {Fassbender} R.,  2005, \mn@doi [\apjl] {10.1086/429801},
  \href {https://ui.adsabs.harvard.edu/abs/2005ApJ...623L..85M} {623, L85}

\bibitem[\protect\citeauthoryear{{Naidu} et~al.,}{{Naidu}
  et~al.}{2022}]{naidu22c}
{Naidu} R.~P.,  et~al., 2022, arXiv e-prints, \href
  {https://ui.adsabs.harvard.edu/abs/2022arXiv220802794N} {p. arXiv:2208.02794}

\bibitem[\protect\citeauthoryear{{Nelson} et~al.,}{{Nelson}
  et~al.}{2023}]{nelson22}
{Nelson} E.~J.,  et~al., 2023, \mn@doi [\apjl] {10.3847/2041-8213/acc1e1},
  \href {https://ui.adsabs.harvard.edu/abs/2023ApJ...948L..18N} {948, L18}

\bibitem[\protect\citeauthoryear{{Nonino} et~al.,}{{Nonino}
  et~al.}{2023}]{nonino22}
{Nonino} M.,  et~al., 2023, \mn@doi [\apjl] {10.3847/2041-8213/ac8e5f}, \href
  {https://ui.adsabs.harvard.edu/abs/2023ApJ...942L..29N} {942, L29}

\bibitem[\protect\citeauthoryear{{Ofek}}{{Ofek}}{2014}]{maat14}
{Ofek} E.~O.,  2014, {MAAT: MATLAB Astronomy and Astrophysics Toolbox},
  Astrophysics Source Code Library, record ascl:1407.005 (\mn@eprint {ascl}
  {1407.005})

\bibitem[\protect\citeauthoryear{{Oguri}, {Takada}, {Okabe}  \&
  {Smith}}{{Oguri} et~al.}{2010}]{oguri10}
{Oguri} M.,  {Takada} M.,  {Okabe} N.,   {Smith} G.~P.,  2010, \mn@doi [\mnras]
  {10.1111/j.1365-2966.2010.16622.x}, \href
  {http://adsabs.harvard.edu/abs/2010MNRAS.405.2215O} {405, 2215}

\bibitem[\protect\citeauthoryear{{Oguri}, {Bayliss}, {Dahle}, {Sharon},
  {Gladders}, {Natarajan}, {Hennawi}  \& {Koester}}{{Oguri}
  et~al.}{2012}]{oguri12}
{Oguri} M.,  {Bayliss} M.~B.,  {Dahle} H.,  {Sharon} K.,  {Gladders} M.~D.,
  {Natarajan} P.,  {Hennawi} J.~F.,   {Koester} B.~P.,  2012, \mn@doi [\mnras]
  {10.1111/j.1365-2966.2011.20248.x}, \href
  {https://ui.adsabs.harvard.edu/abs/2012MNRAS.420.3213O} {420, 3213}

\bibitem[\protect\citeauthoryear{{Okabe}, {Okura}  \& {Futamase}}{{Okabe}
  et~al.}{2010}]{okabe10}
{Okabe} N.,  {Okura} Y.,   {Futamase} T.,  2010, \mn@doi [\apj]
  {10.1088/0004-637X/713/1/291}, \href
  {https://ui.adsabs.harvard.edu/abs/2010ApJ...713..291O} {713, 291}

\bibitem[\protect\citeauthoryear{{Oke} \& {Gunn}}{{Oke} \&
  {Gunn}}{1983}]{oke83}
{Oke} J.~B.,  {Gunn} J.~E.,  1983, \mn@doi [\apj] {10.1086/160817}, \href
  {http://adsabs.harvard.edu/abs/1983ApJ...266..713O} {266, 713}

\bibitem[\protect\citeauthoryear{{Owers}, {Randall}, {Nulsen}, {Couch}, {David}
   \& {Kempner}}{{Owers} et~al.}{2011}]{owers11}
{Owers} M.~S.,  {Randall} S.~W.,  {Nulsen} P.~E.~J.,  {Couch} W.~J.,  {David}
  L.~P.,   {Kempner} J.~C.,  2011, \mn@doi [\apj] {10.1088/0004-637X/728/1/27},
  \href {http://cdsads.u-strasbg.fr/abs/2011ApJ...728...27O} {728, 27}

\bibitem[\protect\citeauthoryear{{Pascale} et~al.,}{{Pascale}
  et~al.}{2022}]{pascale22}
{Pascale} M.,  et~al., 2022, \mn@doi [\apjl] {10.3847/2041-8213/ac9316}, \href
  {https://ui.adsabs.harvard.edu/abs/2022ApJ...938L...6P} {938, L6}

\bibitem[\protect\citeauthoryear{{Price-Whelan} et~al.,}{{Price-Whelan}
  et~al.}{2018}]{astropy18}
{Price-Whelan} A.~M.,  et~al., 2018, \mn@doi [\aj] {10.3847/1538-3881/aabc4f},
  \href {https://ui.adsabs.harvard.edu/#abs/2018AJ....156..123T} {156, 123}

\bibitem[\protect\citeauthoryear{{Priewe}, {Williams}, {Liesenborgs}, {Coe}  \&
  {Rodney}}{{Priewe} et~al.}{2017}]{priewe17}
{Priewe} J.,  {Williams} L. L.~R.,  {Liesenborgs} J.,  {Coe} D.,   {Rodney}
  S.~A.,  2017, \mn@doi [\mnras] {10.1093/mnras/stw2785}, \href
  {https://ui.adsabs.harvard.edu/abs/2017MNRAS.465.1030P} {465, 1030}

\bibitem[\protect\citeauthoryear{{Redlich}, {Bartelmann}, {Waizmann}  \&
  {Fedeli}}{{Redlich} et~al.}{2012}]{redlich12}
{Redlich} M.,  {Bartelmann} M.,  {Waizmann} J.~C.,   {Fedeli} C.,  2012,
  \mn@doi [\aap] {10.1051/0004-6361/201219722}, \href
  {https://ui.adsabs.harvard.edu/abs/2012A&A...547A..66R} {547, A66}

\bibitem[\protect\citeauthoryear{{Richard} et~al.,}{{Richard}
  et~al.}{2014}]{richard14}
{Richard} J.,  et~al., 2014, \mn@doi [\mnras] {10.1093/mnras/stu1395}, \href
  {https://ui.adsabs.harvard.edu/abs/2014MNRAS.444..268R} {444, 268}

\bibitem[\protect\citeauthoryear{{Richard} et~al.,}{{Richard}
  et~al.}{2021}]{richard21}
{Richard} J.,  et~al., 2021, \mn@doi [\aap] {10.1051/0004-6361/202039462},
  \href {https://ui.adsabs.harvard.edu/abs/2021A&A...646A..83R} {646, A83}

\bibitem[\protect\citeauthoryear{{Rieke}, {Kelly}  \& {Horner}}{{Rieke}
  et~al.}{2005}]{rieke05}
{Rieke} M.~J.,  {Kelly} D.,   {Horner} S.,  2005, {Overview of James Webb Space
  Telescope and NIRCam's Role}.
SPIE, pp~1--8, \mn@doi{10.1117/12.615554}

\bibitem[\protect\citeauthoryear{{Rieke} et~al.,}{{Rieke}
  et~al.}{2023}]{rieke23}
{Rieke} M.~J.,  et~al., 2023, \mn@doi [\pasp] {10.1088/1538-3873/acac53}, \href
  {https://ui.adsabs.harvard.edu/abs/2023PASP..135b8001R} {135, 028001}

\bibitem[\protect\citeauthoryear{{Roberts-Borsani} et~al.,}{{Roberts-Borsani}
  et~al.}{2022}]{roberts-borsani22}
{Roberts-Borsani} G.,  et~al., 2022, arXiv e-prints, \href
  {https://ui.adsabs.harvard.edu/abs/2022arXiv221015639R} {p. arXiv:2210.15639}

\bibitem[\protect\citeauthoryear{{Rodney} et~al.,}{{Rodney}
  et~al.}{2015}]{rodney15}
{Rodney} S.~A.,  et~al., 2015, \mn@doi [\apj] {10.1088/0004-637X/811/1/70},
  \href {https://ui.adsabs.harvard.edu/abs/2015ApJ...811...70R} {811, 70}

\bibitem[\protect\citeauthoryear{{Rodney}, {Brammer}, {Pierel}, {Richard},
  {Toft}, {O'Connor}, {Akhshik}  \& {Whitaker}}{{Rodney}
  et~al.}{2021}]{rodney22}
{Rodney} S.~A.,  {Brammer} G.~B.,  {Pierel} J. D.~R.,  {Richard} J.,  {Toft}
  S.,  {O'Connor} K.~F.,  {Akhshik} M.,   {Whitaker} K.~E.,  2021, \mn@doi
  [Nature Astronomy] {10.1038/s41550-021-01450-9}, \href
  {https://ui.adsabs.harvard.edu/abs/2021NatAs...5.1118R} {5, 1118}

\bibitem[\protect\citeauthoryear{{Sereno}, {Jetzer}  \& {Lubini}}{{Sereno}
  et~al.}{2010}]{sereno10}
{Sereno} M.,  {Jetzer} P.,   {Lubini} M.,  2010, \mn@doi [\mnras]
  {10.1111/j.1365-2966.2010.16248.x}, \href
  {https://ui.adsabs.harvard.edu/abs/2010MNRAS.403.2077S} {403, 2077}

\bibitem[\protect\citeauthoryear{{Sharon}, {Gladders}, {Rigby}, {Wuyts},
  {Koester}, {Bayliss}  \& {Barrientos}}{{Sharon} et~al.}{2012}]{sharon12}
{Sharon} K.,  {Gladders} M.~D.,  {Rigby} J.~R.,  {Wuyts} E.,  {Koester} B.~P.,
  {Bayliss} M.~B.,   {Barrientos} L.~F.,  2012, \mn@doi [\apj]
  {10.1088/0004-637X/746/2/161}, \href
  {https://ui.adsabs.harvard.edu/abs/2012ApJ...746..161S} {746, 161}

\bibitem[\protect\citeauthoryear{{Springel} et~al.,}{{Springel}
  et~al.}{2005}]{springel05}
{Springel} V.,  et~al., 2005, \mn@doi [\nat] {10.1038/nature03597}, \href
  {https://ui.adsabs.harvard.edu/abs/2005Natur.435..629S} {435, 629}

\bibitem[\protect\citeauthoryear{{Stanford} et~al.,}{{Stanford}
  et~al.}{2012}]{stanford12}
{Stanford} S.~A.,  et~al., 2012, \mn@doi [\apj] {10.1088/0004-637X/753/2/164},
  \href {https://ui.adsabs.harvard.edu/abs/2012ApJ...753..164S} {753, 164}

\bibitem[\protect\citeauthoryear{{Steinhardt} et~al.,}{{Steinhardt}
  et~al.}{2020}]{steinhardt20}
{Steinhardt} C.~L.,  et~al., 2020, \mn@doi [\apjs] {10.3847/1538-4365/ab75ed},
  \href {https://ui.adsabs.harvard.edu/abs/2020ApJS..247...64S} {247, 64}

\bibitem[\protect\citeauthoryear{{Strait} et~al.,}{{Strait}
  et~al.}{2020}]{strait20}
{Strait} V.,  et~al., 2020, \mn@doi [\apj] {10.3847/1538-4357/ab5daf}, \href
  {https://ui.adsabs.harvard.edu/abs/2020ApJ...888..124S} {888, 124}

\bibitem[\protect\citeauthoryear{{Strait} et~al.,}{{Strait}
  et~al.}{2021}]{strait21}
{Strait} V.,  et~al., 2021, \mn@doi [\apj] {10.3847/1538-4357/abe533}, \href
  {https://ui.adsabs.harvard.edu/abs/2021ApJ...910..135S} {910, 135}

\bibitem[\protect\citeauthoryear{{Torri}, {Meneghetti}, {Bartelmann},
  {Moscardini}, {Rasia}  \& {Tormen}}{{Torri} et~al.}{2004}]{torri04}
{Torri} E.,  {Meneghetti} M.,  {Bartelmann} M.,  {Moscardini} L.,  {Rasia} E.,
   {Tormen} G.,  2004, \mn@doi [\mnras] {10.1111/j.1365-2966.2004.07508.x},
  \href {https://ui.adsabs.harvard.edu/abs/2004MNRAS.349..476T} {349, 476}

\bibitem[\protect\citeauthoryear{{Treu} et~al.,}{{Treu} et~al.}{2015}]{treu15}
{Treu} T.,  et~al., 2015, \mn@doi [\apj] {10.1088/0004-637X/812/2/114}, \href
  {https://ui.adsabs.harvard.edu/abs/2015ApJ...812..114T} {812, 114}

\bibitem[\protect\citeauthoryear{{Treu} et~al.,}{{Treu} et~al.}{2022}]{treu22}
{Treu} T.,  et~al., 2022, \mn@doi [\apj] {10.3847/1538-4357/ac8158}, \href
  {https://ui.adsabs.harvard.edu/abs/2022ApJ...935..110T} {935, 110}

\bibitem[\protect\citeauthoryear{{Tully} \& {Fisher}}{{Tully} \&
  {Fisher}}{1977}]{tully77}
{Tully} R.~B.,  {Fisher} J.~R.,  1977, \aap, \href
  {https://ui.adsabs.harvard.edu/abs/1977A&A....54..661T} {54, 661}

\bibitem[\protect\citeauthoryear{Vale \& Ostriker}{Vale \&
  Ostriker}{2004}]{vale04}
Vale A.,  Ostriker J.~P.,  2004, \mn@doi [Monthly Notices of the Royal
  Astronomical Society] {10.1111/j.1365-2966.2004.08059.x}, 353, 189

\bibitem[\protect\citeauthoryear{{Virtanen} et~al.,}{{Virtanen}
  et~al.}{2020}]{virtanen20}
{Virtanen} P.,  et~al., 2020, \mn@doi [Nature Methods]
  {https://doi.org/10.1038/s41592-019-0686-2}, \href {https://rdcu.be/b08Wh}
  {17, 261}

\bibitem[\protect\citeauthoryear{{Wallington} \& {Narayan}}{{Wallington} \&
  {Narayan}}{1993}]{wallington93}
{Wallington} S.,  {Narayan} R.,  1993, \mn@doi [\apj] {10.1086/172222}, \href
  {https://ui.adsabs.harvard.edu/abs/1993ApJ...403..517W} {403, 517}

\bibitem[\protect\citeauthoryear{{Wang} et~al.,}{{Wang} et~al.}{2015}]{wang15}
{Wang} X.,  et~al., 2015, \mn@doi [\apj] {10.1088/0004-637X/811/1/29}, \href
  {https://ui.adsabs.harvard.edu/abs/2015ApJ...811...29W} {811, 29}

\bibitem[\protect\citeauthoryear{{Wang} et~al.,}{{Wang} et~al.}{2016}]{wang16}
{Wang} T.,  et~al., 2016, \mn@doi [\apj] {10.3847/0004-637X/828/1/56}, \href
  {https://ui.adsabs.harvard.edu/abs/2016ApJ...828...56W} {828, 56}

\bibitem[\protect\citeauthoryear{{Weaver} et~al.,}{{Weaver}
  et~al.}{2023}]{weaver23}
{Weaver} J.~R.,  et~al., 2023, \mn@doi [arXiv e-prints]
  {10.48550/arXiv.2301.02671}, \href
  {https://ui.adsabs.harvard.edu/abs/2023arXiv230102671W} {p. arXiv:2301.02671}

\bibitem[\protect\citeauthoryear{{Welch} et~al.,}{{Welch}
  et~al.}{2022}]{welch22}
{Welch} B.,  et~al., 2022, \mn@doi [\nat] {10.1038/s41586-022-04449-y}, \href
  {https://ui.adsabs.harvard.edu/abs/2022Natur.603..815W} {603, 815}

\bibitem[\protect\citeauthoryear{{Williams} et~al.,}{{Williams}
  et~al.}{2023}]{williams22}
{Williams} H.,  et~al., 2023, \mn@doi [Science] {10.1126/science.adf5307},
  \href {https://ui.adsabs.harvard.edu/abs/2023Sci...380..416W} {380, 416}

\bibitem[\protect\citeauthoryear{{Zavala} et~al.,}{{Zavala}
  et~al.}{2023}]{zavala22}
{Zavala} J.~A.,  et~al., 2023, \mn@doi [\apjl] {10.3847/2041-8213/acacfe},
  \href {https://ui.adsabs.harvard.edu/abs/2023ApJ...943L...9Z} {943, L9}

\bibitem[\protect\citeauthoryear{{Zeldovich}, {Einasto}  \&
  {Shandarin}}{{Zeldovich} et~al.}{1982}]{zeldovich82}
{Zeldovich} I.~B.,  {Einasto} J.,   {Shandarin} S.~F.,  1982, \mn@doi [\nat]
  {10.1038/300407a0}, \href
  {https://ui.adsabs.harvard.edu/abs/1982Natur.300..407Z} {300, 407}

\bibitem[\protect\citeauthoryear{{Zheng} et~al.,}{{Zheng}
  et~al.}{2014}]{zheng14}
{Zheng} W.,  et~al., 2014, \mn@doi [\apj] {10.1088/0004-637X/795/1/93}, \href
  {https://ui.adsabs.harvard.edu/abs/2014ApJ...795...93Z} {795, 93}

\bibitem[\protect\citeauthoryear{{Zitrin}}{{Zitrin}}{2021}]{zitrin21}
{Zitrin} A.,  2021, \mn@doi [\apj] {10.3847/1538-4357/ac0e32}, \href
  {https://ui.adsabs.harvard.edu/abs/2021ApJ...919...54Z} {919, 54}

\bibitem[\protect\citeauthoryear{{Zitrin} et~al.,}{{Zitrin}
  et~al.}{2009}]{zitrin09a}
{Zitrin} A.,  et~al., 2009, \mn@doi [\mnras]
  {10.1111/j.1365-2966.2009.14899.x}, \href
  {https://ui.adsabs.harvard.edu/abs/2009MNRAS.396.1985Z} {396, 1985}

\bibitem[\protect\citeauthoryear{{Zitrin} et~al.,}{{Zitrin}
  et~al.}{2013a}]{zitrin13a}
{Zitrin} A.,  et~al., 2013a, \mn@doi [\apjl] {10.1088/2041-8205/762/2/L30},
  \href {http://adsabs.harvard.edu/abs/2013ApJ...762L..30Z} {762, L30}

\bibitem[\protect\citeauthoryear{{Zitrin}, {Menanteau}, {Hughes}, {Coe},
  {Barrientos}, {Infante}  \& {Mandelbaum}}{{Zitrin} et~al.}{2013b}]{zitrin13b}
{Zitrin} A.,  {Menanteau} F.,  {Hughes} J.~P.,  {Coe} D.,  {Barrientos} L.~F.,
  {Infante} L.,   {Mandelbaum} R.,  2013b, \mn@doi [\apjl]
  {10.1088/2041-8205/770/1/L15}, \href
  {https://ui.adsabs.harvard.edu/abs/2013ApJ...770L..15Z} {770, L15}

\bibitem[\protect\citeauthoryear{{Zitrin} et~al.,}{{Zitrin}
  et~al.}{2014}]{zitrin14}
{Zitrin} A.,  et~al., 2014, \mn@doi [\apjl] {10.1088/2041-8205/793/1/L12},
  \href {https://ui.adsabs.harvard.edu/abs/2014ApJ...793L..12Z} {793, L12}

\bibitem[\protect\citeauthoryear{{Zitrin} et~al.,}{{Zitrin}
  et~al.}{2015}]{zitrin15a}
{Zitrin} A.,  et~al., 2015, \mn@doi [\apj] {10.1088/0004-637X/801/1/44}, \href
  {https://ui.adsabs.harvard.edu/abs/2015ApJ...801...44Z} {801, 44}

\bibitem[\protect\citeauthoryear{{de Lapparent}, {Geller}  \& {Huchra}}{{de
  Lapparent} et~al.}{1986}]{delapparent86}
{de Lapparent} V.,  {Geller} M.~J.,   {Huchra} J.~P.,  1986, \mn@doi [\apjl]
  {10.1086/184625}, \href
  {https://ui.adsabs.harvard.edu/abs/1986ApJ...302L...1D} {302, L1}

\bibitem[\protect\citeauthoryear{{van der Walt}, {Colbert}  \&
  {Varoquaux}}{{van der Walt} et~al.}{2011}]{vanderwalt11}
{van der Walt} S.,  {Colbert} S.~C.,   {Varoquaux} G.,  2011, Computing in
  Science Engineering, 13, 22

\makeatother
\end{thebibliography}



\appendix


\bsp	
\label{lastpage}
\end{document}